\documentclass[showpacs,prl,onecolumn,aps,superscriptaddress,preprintnumbers,nofootinbib]{revtex4}
\usepackage[T1]{fontenc}
\usepackage[latin9]{inputenc}
\usepackage{amsmath,amssymb}
\usepackage{epsfig}
\usepackage{graphicx}
\usepackage{mathrsfs}
\usepackage{amsmath}
\usepackage{amsfonts}

\usepackage{epstopdf}
\usepackage{color}
\def\slashchar#1{\setbox0=\hbox{$#1$}     		
   \dimen0=\wd0                                 	
   \setbox1=\hbox{/} \dimen1=\wd1               	
   \ifdim\dimen0>\dimen1                        	
      \rlap{\hbox to \dimen0{\hfil/\hfil}}      	
      #1                                        	
   \else                                        	
      \rlap{\hbox to \dimen1{\hfil$#1$\hfil}}   	
      /                                         	
   \fi}

\renewcommand{\vec}{\boldsymbol}
\newcommand{\beq}{\begin{equation}}
\newcommand{\eeq}{\end{equation}}
\newcommand{\bea}{\begin{eqnarray}}
\newcommand{\eea}{\end{eqnarray}}
\newcommand{\ba}{\begin{array}}
\newcommand{\ea}{\end{array}}

\def\eq#1{{Eq.~(\ref{#1})}}
\def\fig#1{{Fig.~\ref{#1}}}
\newcommand{\bas}{\bar{\alpha}_S}
\newcommand{\as}{\alpha_S}
\newcommand{\nn}{\nonumber}

\newcommand{\Lb}{\left(}
\newcommand{\Rb}{\right)}
\newcommand{\h}{\frac{1}{2}}
\def\pom{{I\!\!P}}

\begin{document}

\title{ Soft Pomeron in the Colour Glass Condensate approach}
\author{Carlos Contreras}
\email{carlos.contreras@usm.cl}
\affiliation{Departamento de F\'isica, Universidad T\'ecnica Federico Santa Mar\'ia,  Avda. Espa\~na 1680, Casilla 110-V, Valpara\'iso, Chile}
\author{ Eugene ~ Levin}
\email{leving@tauex.tau.ac.il, eugeny.levin@usm.cl}
\affiliation{Departamento de F\'isica, Universidad T\'ecnica Federico Santa Mar\'ia,  Avda. Espa\~na 1680, Casilla 110-V, Valpara\'iso, Chile}
\affiliation{Centro Cient\'ifico-
Tecnol\'ogico de Valpara\'iso, Avda. Espa\~na 1680, Casilla 110-V, Valpara\'iso, Chile}
\affiliation{Department of Particle Physics, School of Physics and Astronomy,
Raymond and Beverly Sackler
 Faculty of Exact Science, Tel Aviv University, Tel Aviv, 69978, Israel}

\author{Michael Sanhueza}
\email{michael.sanhuezar@sansano.usm.cl}
\affiliation{Departamento de F\'isica, Universidad T\'ecnica Federico Santa Mar\'ia,   Avda. Espa\~na 1680, Casilla 110-V, Valpara\'iso, Chile}
\date{\today}

\pacs{13.60.Hb, 12.38.Cy}

\begin{abstract}

      In this paper         we suggest a new approach to the structure of the soft Pomeron:  based on the $t$-channel unitarity,  we expressed the exchange of the soft Pomeron through the interaction of the dipole of  small size of the order of $1/Q_s(Y)$  
  ($Q_s(Y)$ is the saturation momentum)  with the hadrons. 
 Therefore, it  is shown that the typical distances  in soft processes are small $r \sim 1/Q_s\Lb \h Y \Rb $, where  $Y \,=\,ln s$.  The saturation momentum, which determines the energy dependence of the scattering amplitude is  proportional to $ Q^2_s\Lb \h Y \Rb \propto\,\exp\Lb\h \lambda\,Y\Rb$, with $\lambda \approx\,0.2$, and  this behaviour 
  is in perfect agreement with phenomenological Donnachie-Landshoff 
 Pomeron.  We demonstrate that the saturation models could describe the experimental data for $\sigma_{tot}, \sigma_{el},\sigma_{diff} $ and $B_{el}$. Hence our approach is a  good first approximation to start discussion of the soft processes in  CGC approach on the  solid theoretical basis.
\end{abstract}
\maketitle

\vspace{-0.5cm}
\tableofcontents
\section{Introduction}
  We believe that high energy scattering can be described in Reggeon Field Theory (RFT) of Quantum Chromodynamics (QCD), which development  has led to understanding  of many characteristic features
  of the processes at high energies, including the phenomenological application to LHC, RHIC and HERA data during the past three decades. 
  
  The basic ideas of RFT go back to pre-QCD era, when in 1967 V.N. Gribov~\cite{GRIB} proposed his diagram technique, that is based  on  a very general picture and properties of high energy exchanges in a local field theory.   These general ideas were assimilated  to  QCD and  worked out  over the years in many papers \cite{BFKL,LI,LIPREV,GLR,GLR1,MUQI,MUPA,MUDI,LIPFT,NAPE,BART,BKP,MV,MUSA,KOLE,BRN,BRAUN,BK,KOLU,JIMWLK1,JIMWLK2,JIMWLK3, JIMWLK4, JIMWLK5,JIMWLK6,JIMWLK7,JIMWLK8, AKLL,KOLU1,KOLUD,BA05,SMITH,KLW,KLLL}.  However, in spite of much work that has been done, the  theoretical framework of RFT is still incomplete. Actually, we face two problems with RFT: the first is the $s$-channel unitarity for dilute-dense parton system scattering,  which is governed by JIMWLK\footnote{Jalilian-Marian, Iancu,   McLerran, Weigert, Leonidov  and Kovner (JIMWLK) equation~\cite{JIMWLK1,JIMWLK2,JIMWLK3, JIMWLK4, JIMWLK5,JIMWLK6,JIMWLK7,JIMWLK8}}  equation~\cite{KLLL}; and the second one is related to the summation of the BFKL Pomeron loops~\cite{MShoshi,IAN,MUSH,LELU,KLP,LMP,LEPP}. 
  
  Bearing this in mind, we cannot be surprised that RFT is not able to describe the soft interaction of hadrons at high energies. On the other hand, the Colour Glass Condensate 
  (CGC) approach as well as its realization in RFT,  leads to a new  saturation scale (saturation momentum $Q_s(Y)$) which increases at large rapidities ($Y$). It gives us a hope that the soft interactions actually stem from sufficient short distances where we can apply RFT  in QCD.  Phenomenological attempts to describe the soft experimental data, based on these ideas with some additional assumptions, turn out to be  rather successful (see Refs.~\cite{TELAVIV,DURHAM} and references therein).
  
   The main building block of the Gribov Pomeron calculus is the exchange of the soft  Pomeron with the Green's function:
   \beq \label{I1}
   G_\pom\Lb Y, Q_T\Rb\,\,=\,\Lb\,\frac{s}{s_0}\Rb^{\alpha_\pom(Q_T)}\,\,=\,\,e^{ \Lb \Delta \,-\,\,\alpha'_\pom\,Q^2_T\Rb\,Y}
  \eeq 
  where $\alpha_\pom\Lb Q_T\Rb\,\,=\,\,\Delta\,-\,\alpha'_\pom\, Q^2_T$ is the Pomeron trajectory and $Q_T$ is the momentum transferred by the Pomeron.
  
  Our goal in this paper is to build the main   ingredient of the RFT  such as soft Pomeron in the Pomeron calculus,  using the JIMWLK evolution. Our basic   idea can be illustrated using the simple  Pomeron Green's function of \eq{I1}.  One can notice that
  this Green's function has the following factorization property:
  \beq \label{I2}
   G_\pom\Lb Y, Q_T\Rb  \,\,\,=\,\,\,   G_\pom\Lb Y - y , Q_T\Rb \,\, G_\pom\Lb y , Q_T\Rb  
   \eeq
   for any value of $y$. Actually, \eq{I2} follows directly from $t$-channel unitarity~\cite{GRIB} and, therefore, has a general origin and,  hence,  should be held in QCD. In sections II, III and IV we will show that this is a correct expectation and, indeed, we  will generalize \eq{I2} to QCD. It should be noted that this generalization includes the integration over the sizes of dipoles with rapidities $Y$.
  On the other hand, the contribution of the Pomeron to hadron-hadron scattering can be written in the form:
   \beq \label{I3}
   N_\pom\Lb Y,Q_T\Rb\,\,\,=\,\,g^2_h\Lb Q_T\Rb\,   G_\pom\Lb Y, Q_T\Rb    
   \eeq
   and using \eq{I2} we can re-write \eq{I3} as follows:
   \beq \label{I4}
      N_\pom\Lb Y,Q_T\Rb\,\,\,=\,\,   N^h_\pom\Lb Y - y ,Q_T\Rb   \,\, N^h_\pom\Lb  y ,Q_T\Rb 
      ~~~\mbox{with}~~~ N^h_\pom\Lb y' ,Q_T\Rb \,\,=\,\, g_h\Lb Q_T\Rb   \,   G_\pom\Lb y', Q_T\Rb
      \eeq
      
      In section V we will show that \eq{I4} can be generalized to QCD with $ N^h_\pom$, that has the meaning of the dipole scattering amplitude with the hadron.  Such an amplitude can be estimated using the non-linear Balitsky-Kovchegov evolution\cite{BK}.  Using the generalization of \eq{I4} we conclude that the  contribution of the dressed BFKL (Balitsky, Fadin, Kuraev and Lipatov) Pomeron~\cite{BFKL} to hadron-hadron scattering amplitude is proportional to the minimal of two saturation momenta:   $Q^2_s\Lb Y - y\Rb$ and $Q^2_s\Lb y\Rb$. Choosing $y = \h Y$ we obtain the largest contribution, which stems from the shortest distances, providing the best theoretical accuracy in perturbative QCD estimates. Since from high energy phenomenology $Q^2_s\Lb y\Rb \,=\,\exp\Lb \lambda\,Y\Rb$ with $\lambda = 0.2- 0.25 $~\cite{SATMOD0,DKLN}, one can see that 
   we expect the intercept of the dressed BFKL Pomeron will be $\Delta \,=\,0.1-0.125$, which is close to the soft phenomenological Donnachie-Landshoff Pomeron~\cite{DOLA} intercept.  It should be pointed out, that the dressed BFKL Pomeron is quite different from the BFKL Pomeron, which has been derived from perturbative QCD in Ref.~\cite{BFKL}, since in our approach the interactions between perturbative BFKL Pomerons have been taken into account in the triple Pomeron vertex and their vertices of interaction with the hadron.  These interactions result in the fact, that the short distances of about $r \sim 1/Q_s$
   contribute to the soft interaction at high energies.
   Small but not equal to zero $\Delta$ means that the exchange of the dressed BFKL Pomeron violates the Froissart theorem~\cite{FROI}. The interaction between dressed Pomerons as well as their interactions with hadrons, have to be found and to be taken into account to obtain the scattering amplitude  of hadron-hadron interaction.  Such a difficult task is certainly out of scope of this paper and perhaps to solve this problem we will need a new theoretical input both from RFT and from non-perturbtive QCD. In this paper for our estimates of the scale of such contributions  we use the simple eikonal, Glauber  formula~\cite{GLAUBER}, which restores the Froissart theorem.
   
        In section VI we will discuss the dressed Pomeron contribution to diffractive production. In conclusions, we summarize our results and discuss the future prospects.

\section{BFKL Pomeron in  the coordinate representation}
  
  It is well known that the scattering amplitude $ N\Lb Y; \vec{r}, \vec{R}; \vec{Q}_T\Rb$ of the dipole with size $r$ and rapidity $Y \,\gg\,1$ with the dipole of the size $R$ at the rest   
 has the following form   in the leading log(1/x) approximation (LLA)   at high energy (see Refs.~\cite{BFKL,LI,LIPREV,NAPE}:
   \beq \label{GF}
  N^{ \rm BFKL}\Lb Y; \vec{r}, \vec{R}; \vec{Q}_T\Rb\,\,\,=\,\,\frac{ r\,R}{16} \sum_{n=-\infty}^{n = \infty} \int^{\infty}_{-\infty} d \nu \,
  e^{\omega\Lb \nu, n\Rb\, Y\,} \frac{1}{\Lb \nu^2 + \Lb \frac{n -1}{2}\Rb^2\Rb\,\Lb \nu^2 + \Lb \frac{n+1}{2}\Rb^2\Rb} \,\,E^{n,\nu}_Q\Lb  r \Rb \,\,E^{n,-\nu}_Q\Lb R\Rb  \eeq

\begin{figure}[ht]
     \begin{center}
     \includegraphics[width=0.3\textwidth]{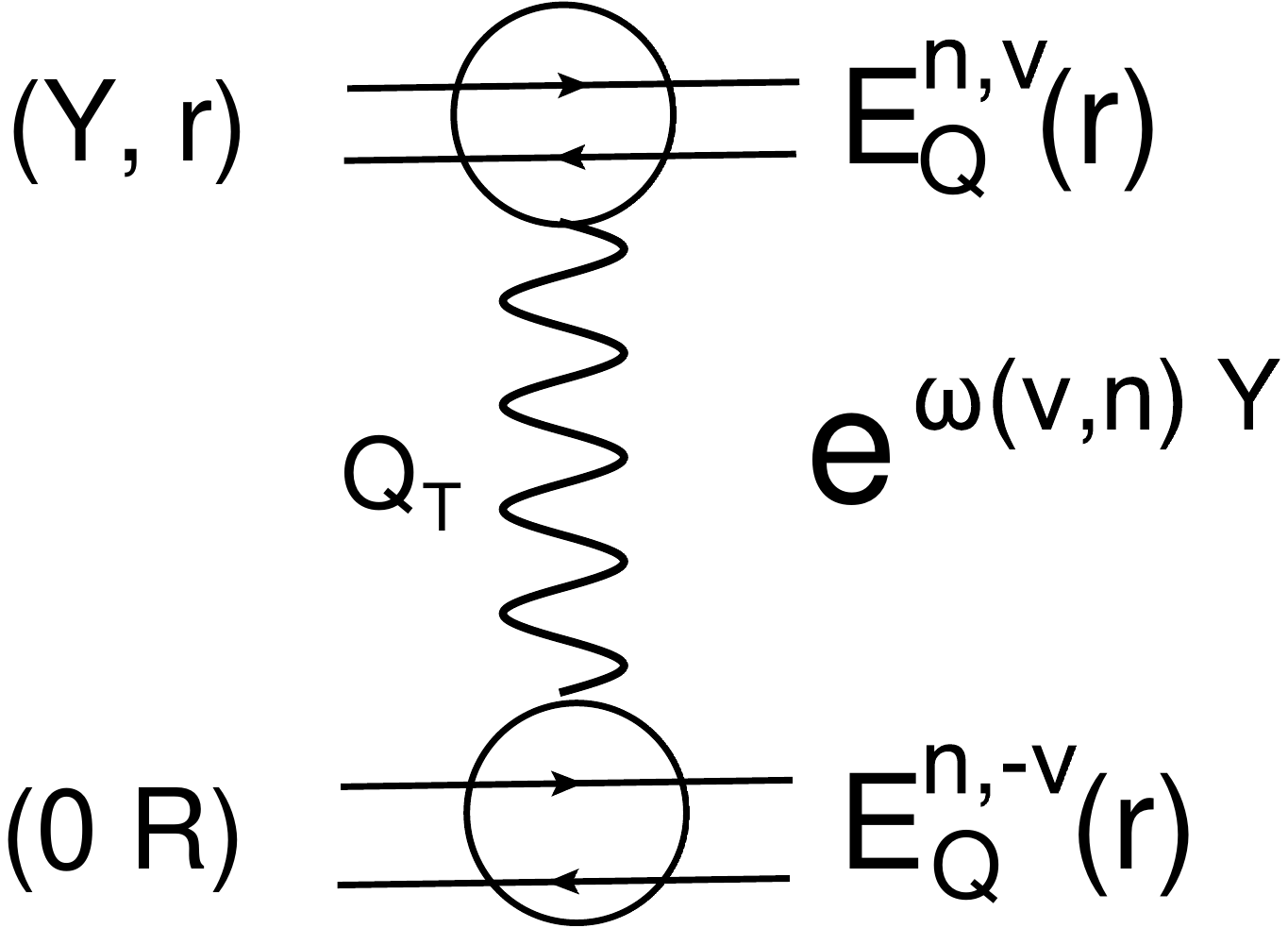} 
     \end{center}    
      \caption{ The general representation of the BFKL Pomeron Green function for the scattering of the dipole with rapidity $Y$ and size $r$ with the dipole with size $R$, which is at the rest. $b$ is the impact parameter of this amplitude.   $Q_T$ is the transverse momentum transferred by the Pomeron.}
\label{pom}
   \end{figure}

  In \eq{GF} $Q_T$ is the transverse momentum that is transferred by the BFKL Pomeron (see \fig{pom}).  One can see that the scattering amplitude can be viewed as the sum of the exchange of the reggeons    whose intercepts are equal to:
   \begin{subequations}  
   \bea 
  \omega\Lb \nu, n\Rb\,\,&=&\,\,2\bas \Bigg( \psi(1) \,\,-\,\,{\rm Re}\Big{\{} \psi\Lb \frac{|n| + 1}{2} \,+\,i\,\nu\Rb \Big{\}}\Bigg);\label{OMEGA}\\
   \omega\Lb \nu, n=0\Rb\,\,&=&\,\,2\bas \Bigg( \psi(1) \,\,-\,\,{\rm Re}\Big{\{} \psi\Lb \frac{ 1}{2} \,+\,i\,\nu\Rb \Big{\}}\Bigg) \,\xrightarrow{\nu\,\ll\,1} \,\, \omega_0 \,+\,D\nu^2 \,=\,4 \ln 2 \bas\,\,+\,\,14 \,\zeta(3)\,\bas \,\nu^2 \label{OMEGA0}
    \eea
    \end{subequations}
      where $\psi(z)$ is the Euler $\psi$-function (see formula {\bf 8.36} of  Ref.~\cite{RY} )  and $\bas \,\,=\,\,\frac{N_c}{\pi} \as$. Generally speaking,   $E^{n,\nu}_Q\Lb r\Rb$ are the Fourier  images of the eigenfunction of the BFKL Hamiltonian in the coordinate space:
\beq \label{EF}
E^{n, \nu}\Lb \rho_{10},\rho_{20}\Rb \,\,=\,\,\Lb - 1\Rb^n\Lb \frac{\rho_{10}\,\rho_{20}}{\rho_{12}}\Rb^{h - \h}\,\,\Lb \frac{\rho^*_{10}\,\rho^*_{20}}{\rho^*_{12}}\Rb^{\tilde{h} - \h}~~~~\mbox{with} ~~~h \,=\,\frac{n}{2} \,-\,i\,\nu; ~~\tilde{h} \,=\,-\,\frac{n}{2} \,-\,i\,\nu;
\eeq
where $\rho_{ik}\,\equiv\,\rho_i \,-\,\rho_k$ are complex transverse coordinates.  They take the form~\cite{LI,LIPREV,NAPE}:

\beq \label{VE}
E^{n, \nu}_Q\Lb r\Rb\,\,\,=\,\,\,\frac{2 \pi^2}{b_{n,\nu}} \,\frac{1}{r}\,\int d z\,d z^* \,e^{\frac{i}{2}\Lb q^*\,z \,\,+\,\, q\,z^*\Rb}\,E^{n, \nu}\Lb z\,+\h \rho, z\,-\,\h \,\rho\Rb
\eeq
where 
\beq \label{BNN}
b_{n,\nu}\,\,=\,\,\frac{2^{4\,i\,\nu}\pi^3}{\h |n| \,-\,i\,\nu}\,\, \frac{ \Gamma\Lb \h |n| \,-\,i\,\nu \,+\,1\Rb \Gamma\Lb \h |n| \,+\,i\,\nu \Rb}{ \Gamma\Lb \h |n| \,+\,i\,\nu \,+\,1\Rb \Gamma\Lb \h |n| \,-\,i\,\nu \Rb};
\eeq

     The explicit form of $E^{n,\nu}_Q\Lb  r \Rb $    
       have been discussed in Refs.~\cite{LI,LIPREV,NAPE} and for $n=0$ they take the forms:
     \beq \label{V}
E^{n=0,\nu}_Q\Lb  r \Rb  =\,\,\Lb Q^2_T\Rb^{i\,\nu} \,2^{- 6 \,i\,\nu} \Gamma^2\Lb 1\,+\,i\,\nu\Rb \Bigg{\{} J_{-i\,\nu}\Lb \frac{q^*\,\rho}{4}\Rb \, J_{-i\,\nu}\Lb \frac{q\,\rho^*}{4}\Rb \,\,-\,\,  J_{i\,\nu}\Lb \frac{q^*\,\rho}{4}\Rb \, J_{i\,\nu}\Lb \frac{q\,\rho^*}{4}\Rb  \Bigg{\}}  
     \eeq      
      In \eq{VE} - \eq{V}  we use the complex number representation for the two-dimensional vectors: 
      $\vec{r} =(x, y) \,\rightarrow\,\,(\rho, \rho^*)$ with $ \rho \,=\,x + i y$ and $ \rho^* \,=\,x\,-\,i\,y$; and
      $\vec{Q}_T =(Q_{T,x} , Q_{T,y} ) \,\rightarrow\,\,(q, q^*)$ with $q \,=\,Q_{T,x} \,+\, i \,Q_{T,y}$ and $q^* \,=\,Q_{T,x}\, -\, i \,Q_{T,y} $ .     

For $Q_T\,\to\,0$ \eq{GF} takes a simple form (see Ref.~\cite{LI} Eq. 32) 
\beq \label{GFQ0}
  N^{ \rm BFKL}\Lb Y; \vec{r}, \vec{R}; \vec{Q}_T\,\to\,0\Rb\,\,\,=\,\,\frac{ r\,R}{8} \sum^{\infty}_{n = - \infty} e^{ i \Lb \varphi \,-\,\psi\Rb \,n} \int^{\infty}_{-\infty} \, d \nu \,
  e^{\omega\Lb \nu, n\Rb\, Y\,} \frac{1}{\Lb \nu^2 + \Lb \frac{n - 1}{2}\Rb^2\Rb\Lb \nu^2 + \Lb \frac{n + 1}{2}\Rb^2\Rb}  \Lb \frac{r^2}{R^2}\Rb^{i\,\nu}
  \eeq
  where $\varphi$ and $\psi$ are angles with the $x$ - axis of $\vec{r}$ and $\vec{R}$, respectively.
  
      Actually \eq{GF} and \eq{GFQ0}  give the scattering amplitude of two dipoles, which satisfies the initial condition:
      \beq \label{IC}
      N^{ \rm BFKL}\Lb Y = 0 ; \vec{r}, \vec{R}; \vec{b}\Rb \,\,=\,\,N^{\rm BA} \Lb \vec{r}, \vec{R}; \vec{b}\Rb   \,\,=\,\,2\pi^2 \ln^2\Lb \frac{
 r^2\,R^2}{\Lb \vec{b}  + \h(\vec{r} - \vec{R})\Rb^2\,\Lb \vec{b} 
 -  \h(\vec{r} - \vec{R})\Rb^2}\Rb      \eeq
      where $N^{\rm BA}$ is the scattering amplitude due to  exchange of two gluons between the dipoles with sizes $r$ and $R$ at the impact parameter $\vec{b}$ (see Refs.~\cite{LI,KOLEB}) .
      
      The scattering amplitudes of \eq{GF} and \eq{GFQ0} can be re-written in more general form:
      
      \beq \label{GFGEN}
     N^{ \rm BFKL}\Lb Y; \vec{r}, \vec{R}; \vec{Q}_T\Rb\,\,\,=\, \sum_{n=-\infty}^{n = \infty} \int^{\infty}_{-\infty} d \nu \,N_{in}\Lb n, \nu\Rb G^{n, \nu}_Q \Lb \vec{r}, \vec{R}\; Y\Rb
     \eeq
     
 where
 \beq \label{GFGEN1}
 G^{n, \nu}_Q \Lb \vec{r}, \vec{R}; Y\Rb\,\,=\,\, e^{\omega\Lb \nu, n\Rb\, Y\,} \,r\,E^{n,\nu}_Q\Lb  r \Rb \,\,R\,E^{n,-\nu}_Q\Lb R\Rb \eeq
 is the Green's function of the BFKL Pomeron with the intercept $  \omega\Lb \nu, n\Rb$ (see \fig{pom}).  $N_{in}\Lb n, \nu\Rb $ have to be found   from the initial condition for the scattering amplitude at $Y=0$.   
\section{ t-channel unitarity for the BFKL Pomeron }
       The BFKL Pomeron is derived in leading logarithmic approximation (LLA) of perturbative QCD using $t$ and $s$ channel unitarity constraints \cite{BFKL,MUDI}. The $s$-channel means that\footnote{For the sake of simplicity we write this constraint at $Q_T=0$.}
       \beq\label{SU}
       {\rm Im}_s N\Lb Y, r,R,Q_T=0\Rb\,\,=\,\,\sum_n \Big{|}N\Lb 2 \to n; \{r_i\}\Rb\Big{|}^2\prod^n_{i=2} d^2 r_i 
       \eeq
  where $N\Lb 2 \to n; \{r_i\}\Rb  $ is the amplitude of production of $n$ dipoles. 
   
\begin{figure}[ht]
     \begin{center}
     \includegraphics[width=0.6\textwidth]{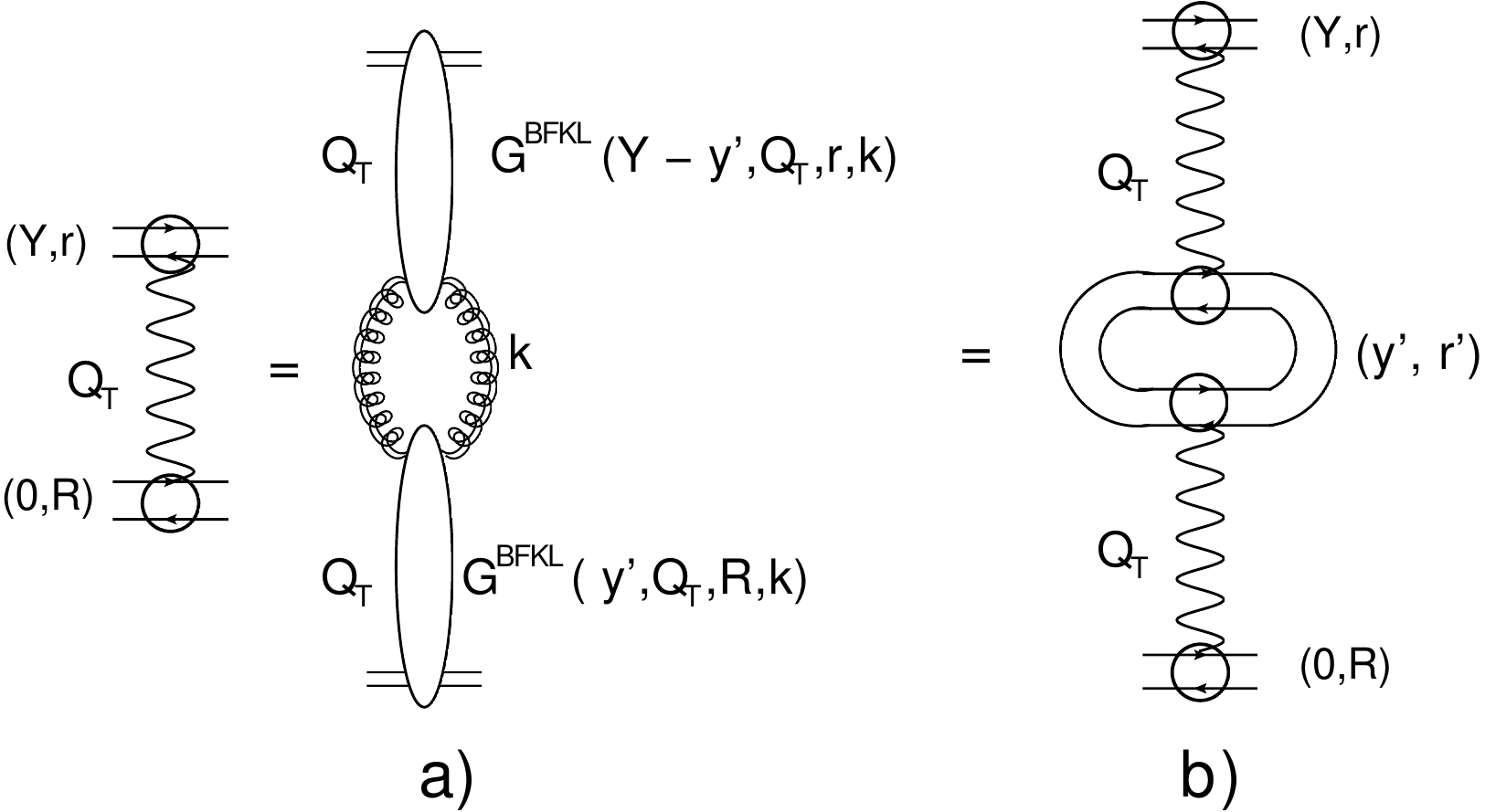} 
     \end{center}    
      \caption{$t$-channel unitarity for the BFKL Pomeron. The double helix lines denote the reggeizied gluons.}
\label{tu}
   \end{figure}

   The BFKL Pomeron satisfies also the $t$-channel unitarity, which in the channel where $t  = - Q^2_T \,>\,0$ is the energy has the same form as \eq{SU}:
   \beq \label{TU}
     {\rm Im}_t N\Lb Y, r,R, Q_T\Rb\,\,=\,\,\sum_n \Big{|}N\Lb 2 \to n; \{k_i\}\Rb\Big{|}^2\prod^n_{i=2} \frac{d^2 k_i }{(2\,\pi)^2}
     \eeq
     where $ N\Lb 2 \to n; \{k_i\}\Rb $ is the amplitude of the production of $n$ gluons with the transverse momenta $k_i$. However, it is shown~\cite{BFKL}, that $t$-channel unitarity, analytically  continued to the $s$-channel, can be re-written as the integration over two reggeized gluons (see \fig{tu}-a) and takes the form\footnote{\eq{TU1} was first written in Refs.~\cite{GLR,MUDI}.}:
   \beq \label{TU1}  
G^{\rm BFKL}\Lb Y, Q_T,  r, R\Rb \,\,=\,\,\int\frac{ d^2 k_T}{(2\,\pi)^2}\, G^{\rm BFKL}\Lb Y - y',Q_T,  r, k_T \Rb   G^{\rm BFKL}\Lb y', Q_T, R, k_T\Rb 
\eeq
where 

\beq \label{GFMR}
 G^{\rm BFKL}\Lb Y - y', Q_T,  r, r'\Rb\,\,\,=\,\, r'^2 \int \frac{d^2 k_T}{(2\,\pi)^2} e^{i \vec{k}_T \cdot\vec{r}'}    G^{\rm BFKL}\Lb Y - y',Q_T,  r, k_T\Rb
 \eeq     
        
    \eq{TU1} can be re-written through $G^{\rm BFKL}\Lb Y - y', r, r', Q_T\Rb$ in the form (see \fig{tu}-b):
  
       \beq \label{TU2}  
G^{\rm BFKL}\Lb Y, Q_T, r, R\Rb \,\,=\,\,\frac{1}{4\,\pi^2}\int\frac{ d^2 r' }{r'^4}\, G^{\rm BFKL}\Lb Y - y', Q_T, r, r'\Rb   G^{\rm BFKL}\Lb y', Q_T,  r', R\Rb 
\eeq     
   The factor $1/4 \pi^2$ in \eq{TU2} we will discuss below. First, let us show that \eq{TU2} holds for the Green's function of \eq{GFGEN1}.  Using the orthogonality of $E_Q^{n, \mu} $ \cite{NAPE}, viz.:
   \beq \label{ORTH}
   \frac{1}{4\,\pi^2}\int \frac{d^2r}{r^2} E^{n,- \nu}_Q\Lb r\Rb\,E^{n, \mu}_Q\Lb r\Rb\, =\,\delta\Lb \nu \,-\,\mu\Rb     
  \eeq
  One can see that   
     \beq \label{TU03}  
G^{n,\nu}_Q\Lb  r, R; Y\Rb \,\,=\,\,\frac{1}{4\,\pi^2}\int\frac{ d^2 r' }{r'^4}\, G^{n, \nu}_Q\Lb  r, r'; Y- y'\Rb   G^{n,\nu}_Q\Lb  r', R; y' \Rb 
\eeq         
   At high energies    the most contribution stems from $n=0$ Green's function  and \eq{TU2} can be  demonstrated directly from \eq{GFQ0} at $Q_T \to 0$:     
   \beq \label{TU3}
 G^{ \rm BFKL}\Lb Y; \vec{r}, \vec{R}; \vec{Q}_T\,\to\,0\Rb\,\,\,=\,\,2\,r\,R \int^{\infty}_{-\infty}  \!\! d \nu  e^{\omega\Lb \nu, 0\Rb\, Y\,} \,\,  \Lb \frac{r^2}{R^2}\Rb^{i\,\nu}
  \eeq  
    \eq{TU2}  can be re-written as follows:
         \bea \label{TU21}  
G^{\rm BFKL}\Lb Y, r, R, Q_T\Rb &=& \frac{1}{( 2\,\pi)^2}\int\frac{ d^2 r' }{r'^4}\, \Bigg\{ 2\,r\,r' \int^{\infty}_{-\infty} \!\!\!\!\! d \nu \,
  e^{\omega\Lb \nu, 0\Rb\, \Lb Y\,-\,y'\Rb} \,\,  \Lb \frac{r^2}{r'^2}\Rb^{i\,\nu} \Bigg\}\,\Bigg\{2\,r'\,R \int^{\infty}_{-\infty}  \!\!\!\!\!  d \nu' \,
  e^{\omega\Lb \nu', 0\Rb\, y'\,} \,\,  \Lb \frac{r'^2}{R^2}\Rb^{i\,\nu'}\Bigg\}\nn\\
  &=&2\, r\,R\,\Bigg\{\int^{\infty}_{-\infty}  \!\!\!\!\!  d \nu \,
  e^{\omega\Lb \nu, 0\Rb\, \Lb Y\,-\,y'\Rb} \,\,  \Lb r^2 \Rb^{i\,\nu} \Bigg\}  \delta(\nu\,-\,\nu')\Bigg\{\int^{\infty}_{-\infty}\!\!\!\!\!  d \nu' \,
  e^{\omega\Lb \nu', 0\Rb\, y'\,} \,\,  \Lb \frac{1}{R^2}\Rb^{i\,\nu'}\Bigg\}\nn\\
 &=&\,\,2\, r\,R\, \int^{\infty}_{-\infty} \!\! \frac{d \nu }{2\,\pi}\,
  e^{\omega\Lb \nu, 0\Rb\, Y\,} \,\,  \Lb \frac{r^2}{R^2}\Rb^{i\,\nu}
\eea
 Note that we checked in \eq{TU21} the numerical factor $1/4\pi^2$, Actually, \eq{TU21} holds for not only $n =  0$ but for all $n$.  
       
     \eq{TU2} can be re-written in the impact parameter representation in the form:
            \beq \label{TUB}  
G^{\rm BFKL}\Lb Y, \vec{r}, \vec{R}, \vec{b}\Rb \,\,=\,\,\frac{1}{4\,\pi^2}\int\frac{ d^2 r' }{r'^4}\,\int d^2 b' \,G^{\rm BFKL}\Lb Y - y', \vec{r}, \vec{r}', \vec{b} - \vec{b}'\Rb \,\,  G^{\rm BFKL}\Lb y',  \vec{r}',\vec{R}, \vec{b}'\Rb 
\eeq     
     \begin{figure}[h]
     \begin{center}
     \includegraphics[width=0.7\textwidth]{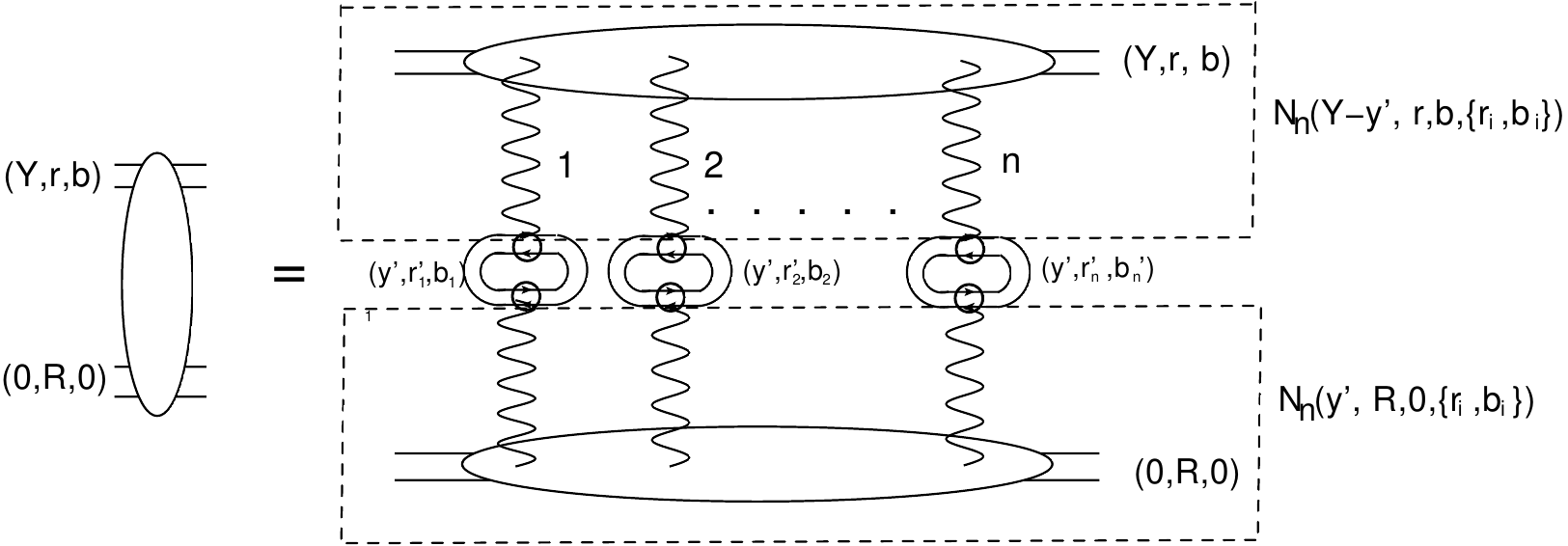} 
     \end{center}    
      \caption{$t$-channel unitarity for a general scattering amplitude in the BFKL Pomeron calculus. }
\label{tug}
   \end{figure}

\section{ t-channel unitarity:  general case }
     
\begin{figure}[ht]
     \begin{center}
     \includegraphics[width=0.25\textwidth]{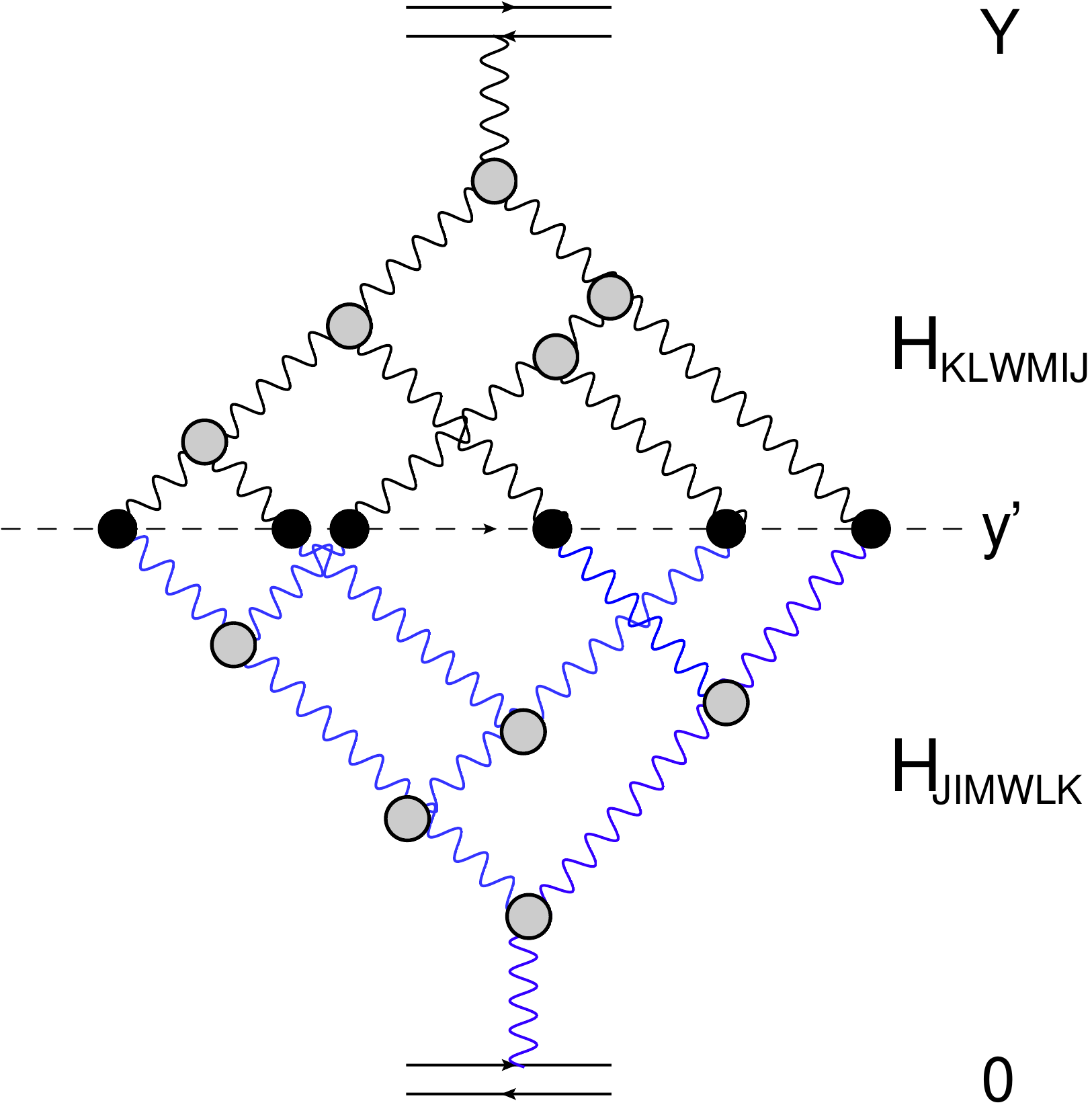} 
     \end{center}    
      \caption{The BFKL cascades ,  which are described by $H_{\mbox{\tiny JIMWLK}}$ and by 
     $H_{\mbox{\tiny KLWMIJ }}$ (see Ref.~\cite{AKLL}).     The wavy lines denote the BFKL Pomerons. The gray circles are the triple Pomeron vertex while
the black circles denote  $ \frac{1}{4 \pi^2} \int d^2 r_i\,d^2 b_i \frac{1}{r^4_i}$.}
\label{lalo}
   \end{figure}

     The t-channel unitarity constraints for the dipole-dipole amplitude can be re-written in a general form in the framework of the BFKL Pomeron calculus  using \eq{TUB} (see \fig{tug}):
     \beq \label{TUGE}
 N\Lb Y, \vec{r}, \vec{R}, \vec{b}\Rb   \,\,\,=\,\,\,\sum^{\infty}_{n=1}\frac{(-1)^{n-1}}{n!}  \int \frac{d^2 r_i\,d^2 b_i}{4\,\pi^2} \,\frac{1}{r^4_i} N_n\Lb Y- y',  \vec{r}, \vec{b}; \{ \vec{r}_i,  \vec{b}_i\}\Rb\,\, N_n\Lb y',  \vec{R}, \vec{0}; \{ \vec{r}_i,  \vec{b}_i\}\Rb 
 \eeq
  where $N_n\Lb Y- y',  \vec{r}, \vec{b}; \{ \vec{r}_i,  \vec{b}_i\}\Rb$ is the amplitude of the production of $n$ BFKL Pomerons each of which produces the dipole with  size $r_i$ at the impact parameters $b_i$. $Y - y'$ is the rapidity between the initial dipole $r$ and produced dipoles $r_i$.

   \eq{TUGE} is a modification of the MPSI\footnote{Mueller, Patel, Salam and Iancu approach.} approach~\cite{MUPA,MPSI} in which we integrated over the sizes of the
    dipoles in  the dipole-dipole scattering amplitudes  at low energies using the properties of the BFKL Pomeron.
    This equation can be useful in the case if we know the amplitudes $N_n$. For example in Ref.~\cite{AKLL} it shown that in the kinematic region $Y - y' \,\leq\, y_{\rm max}$ and $y'\, \leq \, y_{\rm max}$ ($y_{\rm max} = \frac{1}{\omega_0}\ln\Lb\frac{1}{\bas^2}\Rb$) $N_n$ are given by the Balitsky-Kovchegov cascade (see \fig{lalo}). In this case \eq{TUGE} allows us to sum the large Pomeron loops as it is shown in \fig{lalo}. \eq{TUGE} can be re-written in this case in the following form~\cite{KO1,LE1}:
\beq  \label{TUGE1}
N\Lb Y, r,R,b\Rb\,\,= \,\,\sum^\infty_{n=1}\frac{ (-1)^{n+1} }{n!} \int \prod^n_{i=1}\frac{1}{4 \,\pi^2}  \frac{d^2 r_i}{r^4_i}\, d^2 b_i\, \,  \frac{ \delta}{\delta u_i} Z\Lb Y - y'; \{ u_i\}\Rb\Big{|}_{u_i=1} \,\, \frac{ \delta}{\delta u'_i} Z\Lb y'; \{ u'_i\}\Rb\Big{|}_{u'_i=1}
 \eeq
   where the generating functional $Z\Lb Y  \{ u_i\}\Rb$ has been discussed in Refs. \cite{ MUDI,LELU}. 
          
In the next section we will give another example of using \eq{TUGE}.

\section{ Dressed BFKL Pomeron in proton-proton scattering }
\subsection{ The master equation }
\begin{figure}[ht]
     \begin{center}
     \includegraphics[width=0.5\textwidth]{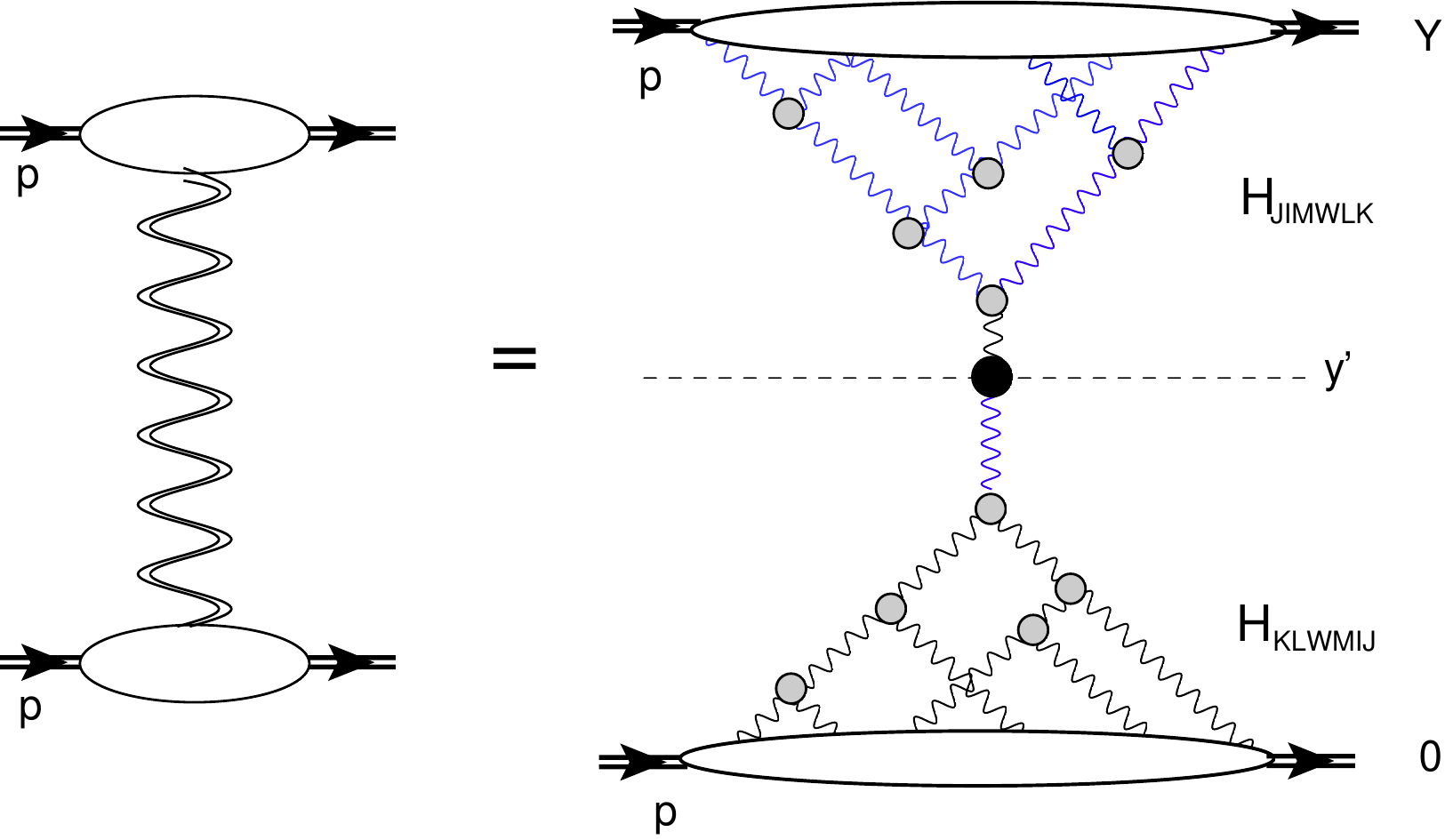} 
     \end{center}    
      \caption{The contribution of the dressed Pomeron to the proton-proton scattering.
     The Balitsky - Kovchegov cascades  are described by $H_{\mbox{\tiny JIMWLK}}$ and by 
     $H_{\mbox{\tiny KLWMIJ }}$ (see Ref.~\cite{AKLL}).     The wavy lines denote the BFKL Pomerons. The gray circles are the triple Pomeron vertex while
the black circle denotes  $ \frac{1}{4 \pi^2} \int d^2 r\,d^2 b_i \frac{1}{r^4}$. The double wavy line describes the dressed Pomeron.}
\label{drpom}
   \end{figure}

    Our main idea is to use \eq{TUGE} to estimate the proton-proton scattering. We believe that for real estimates we need to find how to sum all Pomeron diagrams including summing of the Pomeron loops.  In spite  of some progress in this direction~\cite{MShoshi,IAN,MUSH,LELU,KLP,LMP,LEPP} we are still far away from the solid theoretical approach both for dilute-dilute parton system scattering  and for dense-dense system interaction.  The example of the first one is the hadron-hadron collisions at high energies while for the second is the nucleus-nucleus scattering.  In this paper we wish to realize a more restricted goal: to build the first approximation to hadron-hadron and/or nucleus-nucleus collisions. We propose  the dressed Pomeron contribution, which is shown in \fig{drpom}, as the  first approximation. In other words, we wish to introduce not the exchange of the BFKL Pomeron as the first approximation but we suggest to sum all Pomeron diagrams that contribute to the vertex for interaction of the BFKL Pomeron with the hadron (see \fig{drpom}). 
    
    We can see that the interaction of the BFKL Pomeron with the proton is known from the DIS data and we have  numerous attempts to describe  this interaction using the Balitsky-Kovchegov  parton  cascade~\cite{SATMOD0,SATMOD1,SATMOD2,IIM,SATMOD3,SATMOD4,SATMOD5,SATMOD6,SATMOD7,SATMOD8,SATMOD9,SATMOD10, SATMOD11,SATMOD12,SATMOD13,SATMOD14}. Therefore, we can develop a model for the vertex.  
        
    Our master equation is shown in \fig{drpom} and  has a simple form:    
    \beq \label{ME}
    N^p_p\Lb Y, b \Rb\,\,\,=\,\,\,\frac{1}{4\,\pi^2} \int 
    \frac{d^2 r'\,d^2 b'}{ r'^4} \,N_p\Lb Y- y' , \vec{r}', \vec{b } - \vec{b}'\Rb\,\,N_p\Lb  y' , \vec{r}', \ \vec{b}'\Rb
    \eeq
    where $N_p$ is the amplitude that can be found from the DIS since all observable in these processes can be expressed through the following amplitudes~\cite{KOLEB}:
    \beq\label{FORMULA}
N\Lb Q, Y; b\Rb \,\,=\,\,\int \frac{d^2 r}{4\,\pi} \int^1_0 d z \,|\Psi_{\gamma^*}\Lb Q, r, z\Rb|^2 \,N_p\Lb r, Y; b\Rb
\eeq      

Note, that the wave functions are known  at least at large values of $Q$.  One can see that integral over $r$ in \eq{ME} converges both at $r\,\to\,0$ and at  large $r \,\to\,\infty$. Indeed,
 at $r'\,\to\,0$  $N_p(Y, r) \,\propto\,\,r^2$ 
 leading to the final integral in the region of small $r$. At large distances $N_p(Y, r) \,\to\,\,1$ and, therefore, the integral is rapidly converges at large distances. 
 
 Hence we expect that the typical $r'$ is about of $1/Q_s$, where $Q_s$ is the saturation scale. Bearing this in mind we expect that the  dressed Pomeron will behave as $Q^2_s\Lb \h Y\Rb$ (for $ y'=\h Y$). Therefore, the dressed Pomeron has the power-like behaviour with intercept $\h \lambda $ if $Q^2_s(Y) \propto \exp\Lb \lambda\,Y\Rb$. Since the phenomenological value of $\lambda =0.2 - 0.3$ we see that the value of the intercept is about $0.1-0.15$ in a good agreement with high energy phenomenology.\footnote{We will discuss this behaviour below in more details.} It should be stressed that this estimate demonstrates that the typical values of $r'$ is rather small ($r' \sim 1/Q_s$) . Therefore, we can safely apply  CGC approach for these calculations and, hence, \eq{ME} gives for the first time  an estimate for soft Pomeron on the solid theoretical basis.  One can see that this estimate of the typical distances is valid for the general \eq{TUGE}, making the approach theoretically very attractive.
        
\subsection{ The simple model for DIS }

    For better understanding of \eq{ME} we model the scattering amplitude $N_p\Lb  y' , \vec{r}', \ \vec{b}\Rb$
 in the following way:
    
    \beq \label{MOD}
N_p\Lb  y , \vec{r}, \ \vec{b}\Rb\,\,=\,\,a\,\Bigg(1\,\,-\,\,\exp\Lb -\tau^{\bar{\gamma}}\,e^{-\frac{b^2}{B}} \Rb\Bigg)\,\,+\,\,a \frac{ \tau^{\bar{\gamma}}e^{-\frac{b^2}{B}}}{1\,\,+\,\,\tau^{\bar{\gamma}}e^{-\frac{b^2}{B}}} \eeq    
\eq{MOD} leads to the scattering amplitude $N_p\Lb  y' , \vec{r}', \ \vec{b}\Rb\,\,=\,\,\tau^{\bar{\gamma}}\,e^{-\frac{b^2}{B}}$ for $\tau\,= r^2\,Q^2_0 \exp\Lb \lambda y\Rb\ \,\ll\,1$. Comparing this behavior with the scattering amplitude in the vicinity of the saturation scale \cite{MUT} we obtain that $\tau^{\bar{\gamma}}\, \,=\,\,N_0 \Lb r^2\,Q^2_s\Lb y, b=0\Rb\Rb^{\bar{\gamma}}$  with
$ Q^2_s\Lb y, b\Rb\,\,=\,\, Q^2_0\,\exp\Lb \lambda\,y\Rb \,e^{-\frac{b^2}{\bar{\gamma}\,B}}$. For $\tau\,>\,1$ \eq{MOD} with a=0.65  gives the good parameterization of the solution to the non-linear Balitsky-Kovchegov (BK) equation for the leading twist \cite{LEPP}.
For the total cross section \eq{ME} takes the form:
\beq \label{XSPP}
\sigma^{pp}_{\rm tot}\,\,=\,\,\frac{2}{ 4 \,\pi^2} \int d^2 r\,\Bigg( \int d^2 b\, N_p\Lb  y', \vec{r}, \ \vec{b}\Rb\Bigg)^2/r^4
\eeq
The integral over $b$ can be taken explicitly, viz.:
\beq \label{XSPP1}
\int d^2 b\, N_p\Lb  y , \vec{r}, \ \vec{b}\Rb\,\,=\,\pi\,B\,\Bigg((1 - a)  \ln (\tau^{\bar{\gamma}}+1)+ a (\ln (\tau^{\bar{\gamma}})+\Gamma (0,\tau^{\bar{\gamma}}) )\Bigg)
\eeq
Using \eq{XSPP1} we can re-write \eq{XSPP}  as follows:
\beq \label{XSPP2}
\sigma^{pp}_{\rm tot}\,\,=\,\,\frac{\pi }{ 2}\, B^2 \, N_0^{\frac{1}{\bar{\gamma}}  }\,Q^2_0 \,e^{\h \lambda  \,y}
 \int d \tau\,\Bigg( (1 - a)  \ln (\tau^{\bar{\gamma}}+1)+ a (\ln (\tau^{\bar{\gamma}})+\Gamma (0,\tau^{\bar{\gamma}}) )  \Bigg)^2/\tau^2
\eeq
For $a=0.65$ and $\bar{\gamma} = 0.63$, which stems from the leading order estimates, the integral over $\tau$ is equal to 4.96.
 Hence  we have for the cross section:
  
\beq \label{XSPP3}
\sigma^{pp}_{\rm tot}\,\,=\,\,\frac{4.96\,\pi}{ 2}\, B^2 \, N_0^{\frac{1}{\bar{\gamma}}  }\,Q^2_0 \,e^{\h \lambda  \,y}
\eeq
The values of $\lambda,  N_0, B$ and $Q^2_0$ have been estimated in the variety of  models~\cite{SATMOD1,SATMOD2,IIM,SATMOD3,SATMOD4,SATMOD5,SATMOD6,SATMOD7,SATMOD8,SATMOD9,SATMOD10,SATMOD11,SATMOD12,SATMOD13,SATMOD14,CLP,CLMP,DIMST,CLS}
which describe the experimental data on DIS from HERA. These models lead to
 $B = 5.5\, GeV^{-2}$, which  can be fixed from the production of $J/\Psi$ meson in DIS; to $N_0  = 0.23 - 0.34$ and  of $\lambda = 0.2 - 0.25$. In the most models $Q^2_0 \approx 0.2\, GeV^2$. Using these values for parameters we have $\sigma^{pp}_{\rm tot}\,=\,39\,mb$ instead of   the experimental value of $\sigma^{pp}_{\rm tot}\,=\,62\,mb$  at $W = 540\,GeV$.  However, the saturation model in the next-to-leading order (see Ref.~\cite{CLMP} for example) lead to larger values of $Q^2_0$. 

\begin{figure}[ht]
     \centering
     \includegraphics[width=0.4\textwidth]{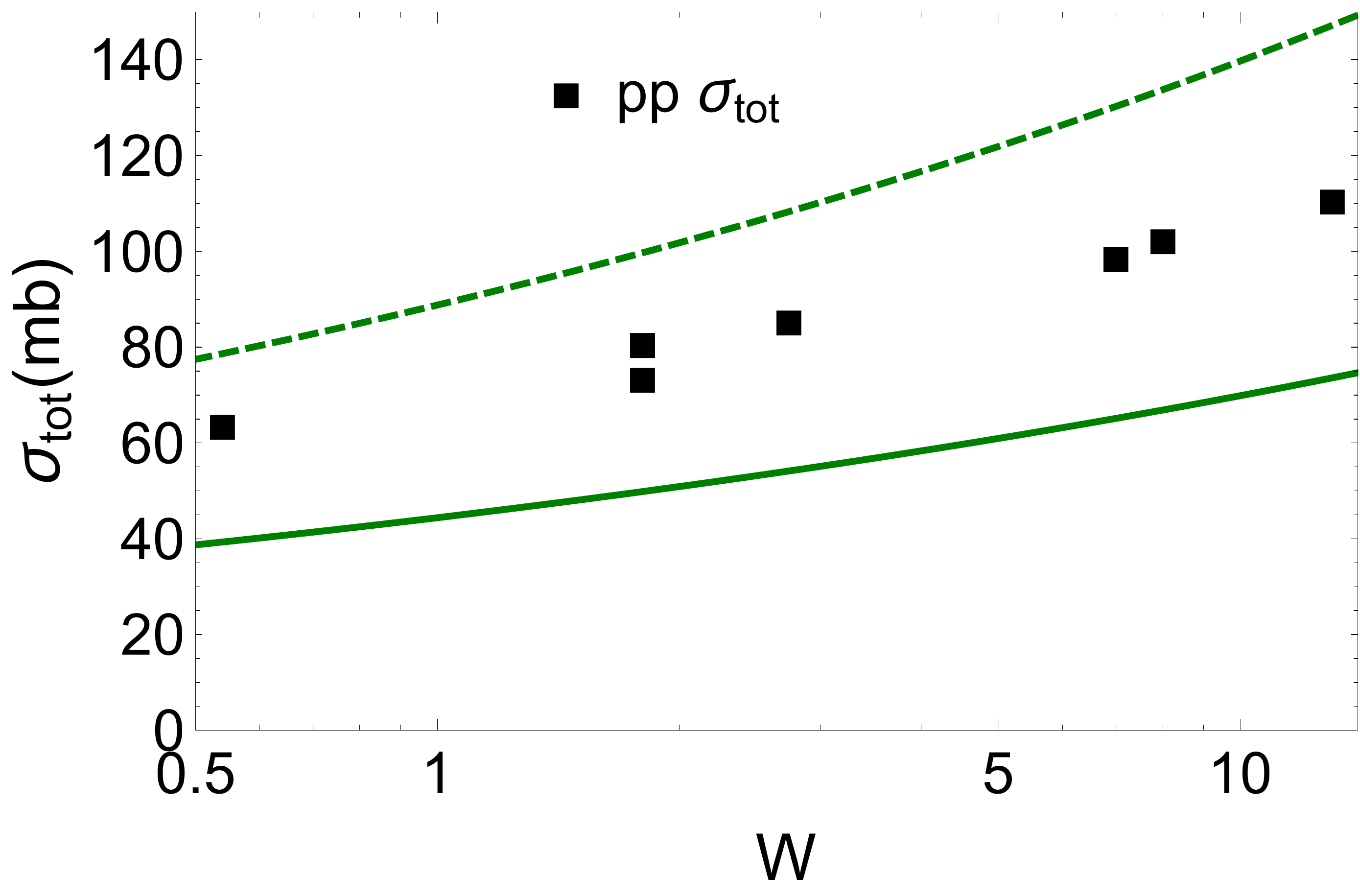}                  
      \caption{   $\sigma_{\rm tot}$ versus W. The  curves  are calculated, using  \eq{XSPP}-\eq{XSPP3}.The solid curve corresponds to $Q^2_0 = 0.2 \,GeV^2$ while the dashed one is for $Q^2_0 = 0.4 \,GeV^2$.  The data are taken from Refs.~\cite{PDG,TOTEM}.   $\lambda = 0.196$, $N_0 = 0.3$,$B = 11 \,GeV^{-2}$.}
\label{xsmod}
   \end{figure}

  In \fig{xsmod} we plot the values for the total cross section for proton-proton scattering (solid line) that come from \eq{MOD} for two values of $Q^2_0 = 0.2\, GeV^2
  $ and $Q^2_0 = 0.4\, GeV^2$.
  
\begin{figure}[ht]
     \begin{center}
     \includegraphics[width=0.5\textwidth]{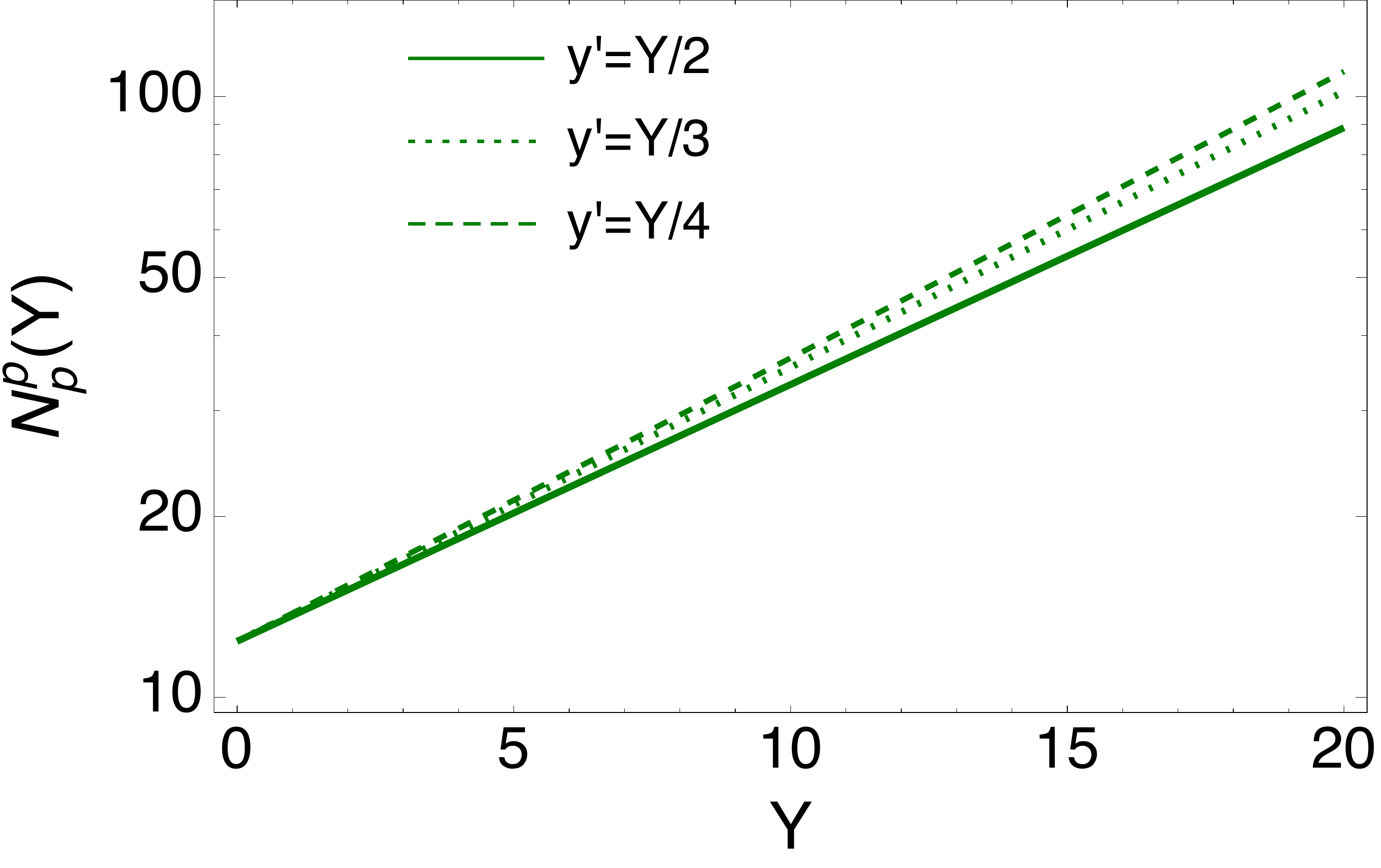} 
     \end{center}    
      \caption{The dressed Pomeron  at $b=0$  versus $Y$ at different values of $y'$. The parameters of \eq{MOD} are taken from Ref.~\cite{CLP}: $N_0$ = 0.34, $\lambda = 0.195$,$Q^2_0=0.145\,GeV^2$ and $ m = 0.75 \,GeV$. The saturation scale is parameterized as $Q_s\Lb Y,b\Rb\,\,=\,\,Q^2_0 \exp\Lb \lambda\, Y\Rb\,S\Lb b\Rb$ with $S\Lb b\Rb\,=\,\Lb m b \, K_1\Lb m \,b\Rb\Rb^{\frac{1}{\bar{\gamma}}}$.
     }
\label{drpvy}
   \end{figure}

         In \fig{drpvy} we show  the dependence of the  dressed Pomeron of \eq{MOD} on $Y$  and $y'$ using this model.  
    From \fig{drpvy} one can see that the contribution of the dressed Pomeron depends on the choice of $y'$.  However, this dependence is not very steep.
As it is shown in Ref.~\cite{MPSI}  the minimal corrections appear at $y'=\h Y$, which we will use in our further estimates.

     It should be stressed that \fig{xsmod} is the first estimates of the cross section  for the soft process that has been made in CGC approach  on solid theoretical ground. We will present the more reliable estimates based on \eq{ME} without using the simplified models. However, these first estimates show us that the approach with the dressed Pomeron can be rather useful.   The simple model led to the cross section which describes the energy behavior of the experimental data. The values of the parameters have  large dispersions,  but for the first estimate we believe, that  we will be able to obtain a good agreement with the data of the cross section values. However, we cannot reproduce the values and energy dependence of $\sigma_{\rm el}$ and $B_{\rm el}$ from \eq{OBSEL} and \eq{OBSBEL}. We will come back to this problem after making more reliable estimates  beyond  the simple model.
     
\subsection{ Realistic estimates }
  
     At first sight for the amplitude  $N_p\Lb  y' , \vec{r}', \ \vec{b}'  \Rb$ in \eq{ME} we can use the non-linear BK equation~\cite{BK}.  However, since  the CGC approach suffers the severe theoretical problem of  violating the Froissart theorem at large impact parameters ($b$)~\cite{KW1,KW2,KW3,FIIM},  we have to build models  which  give the scattering amplitude  with exponential  decrease   at large $b$. All these models use
   the theoretically solid  behaviour of the scattering amplitude in the vicinity of the saturation scale  ($\tau \,=\,r^2 \,Q^2_s\Lb y, b\Rb \,\,\sim\,\,1$)~\cite{MUT}:
   \beq \label{RE1}
    N_p\Lb y, r, b \Rb\,\,=\,\, N_0 \Lb r^2 \,Q^2_s\Lb y, b\Rb\Rb^{\bar \gamma}
     \eeq
     with $\bar{\gamma} = 0.63$ in the leading order of perturbative QCD.

\begin{figure}[ht]
     \begin{center}
     \includegraphics[width=0.8\textwidth]{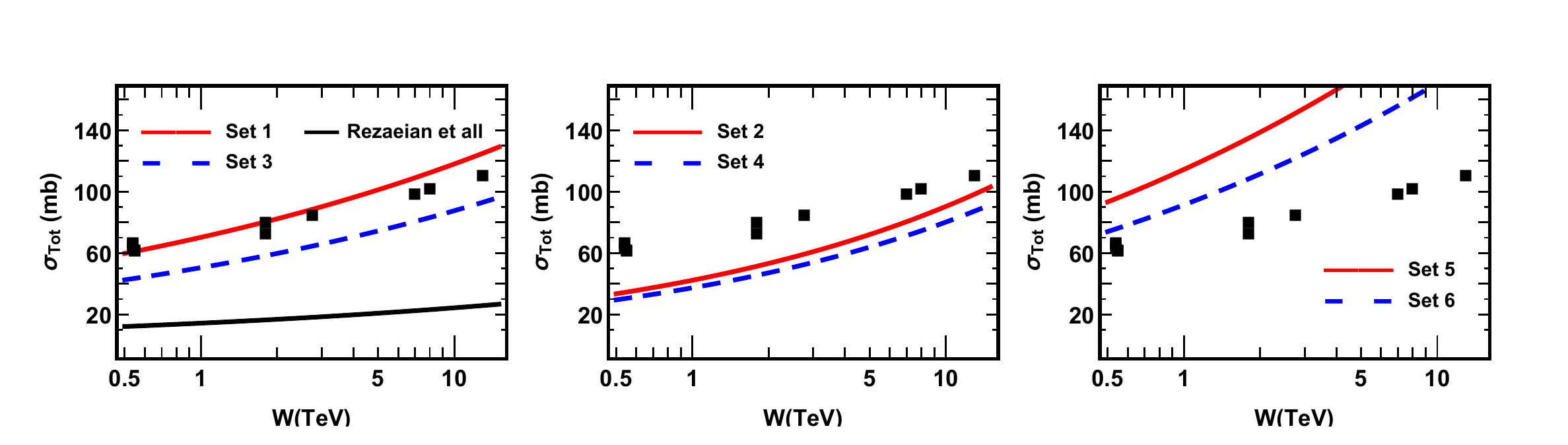} 
     \end{center}    
      \caption{ $\sigma_{\rm tot}$ versus W from \eq{XSPP} for all sets of Ref.~\cite{CLS}.
      The solid black line corresponds to the saturation model of Ref.~\cite{SATMOD14}.
       }
\label{set13}
   \end{figure}

    However, for the $b$-dependence of the saturation scale   the phenomenological  $\exp\Lb - \mu \,b\Rb$ or $\exp\Lb - b^2/B\Rb$ behaviour  is taken instead of the power-like decrease, that follows from the BK equation.  For $\tau\,>\,1$  it is assumed the geometric scaling behaviour of the scattering amplitude~\cite{BALE,IIML,SGBK}, which leads to $N_p = N_p(\tau)$.  We need to use BK equation to find this function.  However, only in Refs.~\cite{CLP,CLMP,CLS} such procedure has been developed. In other models  the rough approximation to the BK equation has been applied. For realistic estimates we chose the model of Ref.~\cite{CLS} , which includes all theoretical ingredients  from the CGC approach (see Ref.~\cite{CLMS}) and introduces the exponential decrease of the saturation scale with $b$ which follows from the Froissart theorem~\cite{FROI}. In \fig{set13} we plot our estimates from \eq{XSPP} for all sets of parameters of Ref.~\cite{CLS}, which demonstrates that the values of the cross sections can be close to the experimental ones. One can see that sets 1 and 3 describe the experimental data while all others  leads to the cross section,  which larger or smaller than the experimental one. Such large dispersion of the estimates is mostly  related to the energy dependence of the saturation scale, which leads to  different values of the typical distances in the integral over $r$ in \eq{XSPP}.

 The large differences between the estimates of the model of Ref.~\cite{CLS} and of Ref.~\cite{SATMOD14} and/or \eq{XSPP3}  stems from the fact that the value of $Q^2_0$ in Ref.~\cite{CLS} is about 1 \,$GeV^2$ being  almost 3-5 times larger that in \fig{xsmod}.
    
It should be noted that \eq{XSPP} cannot describe the energy dependence of the slope for differential elastic cross section as well as the value of $\sigma_{el}$. 
Bearing this in mind we made the estimates for the shadowing corrections  using the eikonal formula:

     \beq \label{EIK}
     A^{pp}\Lb Y, b\Rb\,\,=\,\,i\, \Bigg(1\,\,-\, \,\exp\Lb -\, N^{p}_p\Lb Y, b\Rb\Rb\Bigg)
     \eeq
     
 The observables can be expressed through the amplitude of \eq{EIK} in the following form:
 \begin{subequations}  
   \bea      
     \sigma_{tot}\Lb Y\Rb\,\,&=& \,\,\,2 \int d^2 b\, {\rm Im}A^{pp}\Lb Y, b\Rb ;\label{OBST}\\
       \sigma_{el}\Lb Y\Rb\,\,&=& \,\,\, \int d^2 b \, |A^{pp}\Lb Y, b\Rb|^2 ;\label{OBSEL}\\   
     B_{el} \,\,&=& \h \int b^2 d^2 b\, {\rm Im}A^{pp}\Lb Y, b\Rb\Bigg{/} \int d^2 b \,{\rm Im}A^{pp}\Lb Y, b\Rb; \label{OBSBEL}
     \eea
     \end{subequations}     
     
\begin{figure}[ht]
     \begin{center}
     \begin{tabular}{c}
     \includegraphics[width=0.8\textwidth]{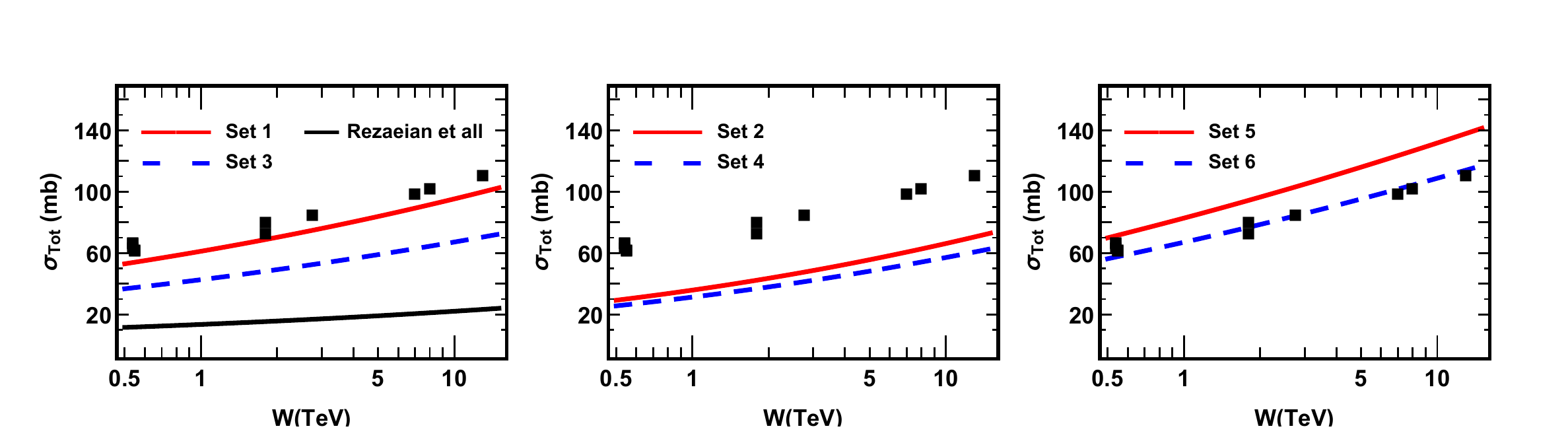} \\
      \includegraphics[width=0.8\textwidth]{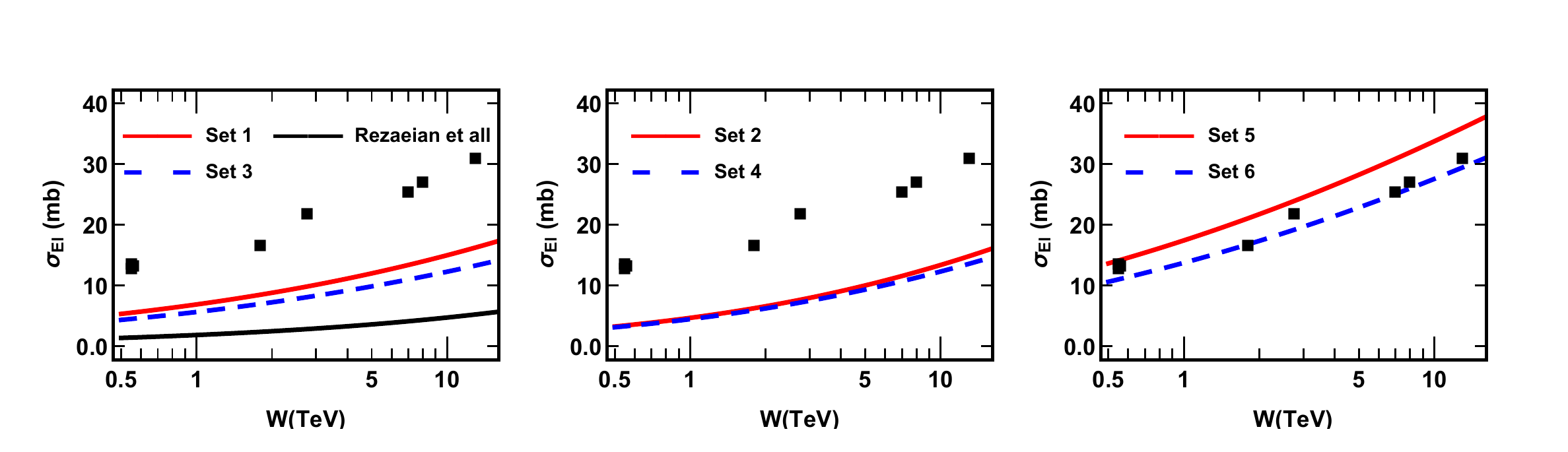} \\  
        \includegraphics[width=0.8\textwidth]{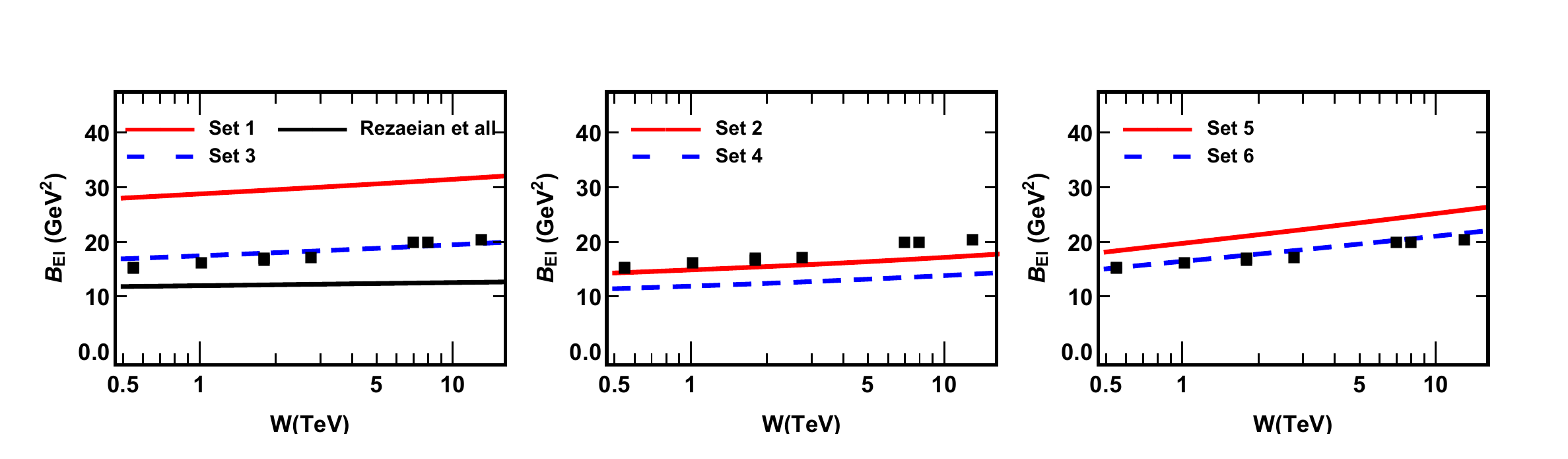} \\   
        \end{tabular}   
         \end{center}    
      \caption{ Comparison of $\sigma_{tot}$, $\sigma_{el}$ and $B_{el}$ with the experimental data for all sets of parameterization of Ref.~\cite{CLS}. The solid black line corresponds to the saturation model of Ref.~\cite{SATMOD14}. The data are taken from Refs.~\cite{PDG,TOTEM}.
       }
\label{setall}
   \end{figure}

     In \fig{setall} we compare our estimates, using \eq{EIK} - \eq{OBSBEL} 
      for all six sets of  parameterization of Ref.~\cite{CLS}. Even a brief sight at \fig{setall} shows  the wide spreading of the values for the observables. This large dispersion of the predictions supports the idea that the DIS data is not enough for fixing the parameters of the models. On the other hand,  one can conclude that we are able to describe both the soft experimental data and DIS. In \fig{setall} sets 5 and 6 describe the data quite well.  It is interesting to note that both these sets    
    introduce the shrinkage of the diffraction peak due to the energy dependence of the impact parameter distribution  for the saturation scale (see, for example, Refs.~\cite{LEPION,GOLEB}).  
    
    It worthwhile mentioning that \eq{EIK} is written as the example of possible shadowing corrections just for understanding the scale of the effect.  As has been discussed in the introduction  the  theoretical approach to these correction is still in the embryonic stage. However,  applying the t-channel unitarity in its general form (see \fig{tug})  we see that all Pomerons (dipoles at   rapidity $y'$) enter at the same typical sizes $r = 1/Q_s$ . Hence, we can try to treat the shadowing corrections in the toy model: the QCD approach in which all dipoles have the same size~\cite{MUDI,LELU1,LELU2,MUSA,BIT,AKLL}.

\section{Diffraction dissociation in the region of large masses}
\begin{figure}[ht]
     \begin{center}
     \includegraphics[width=0.7\textwidth]{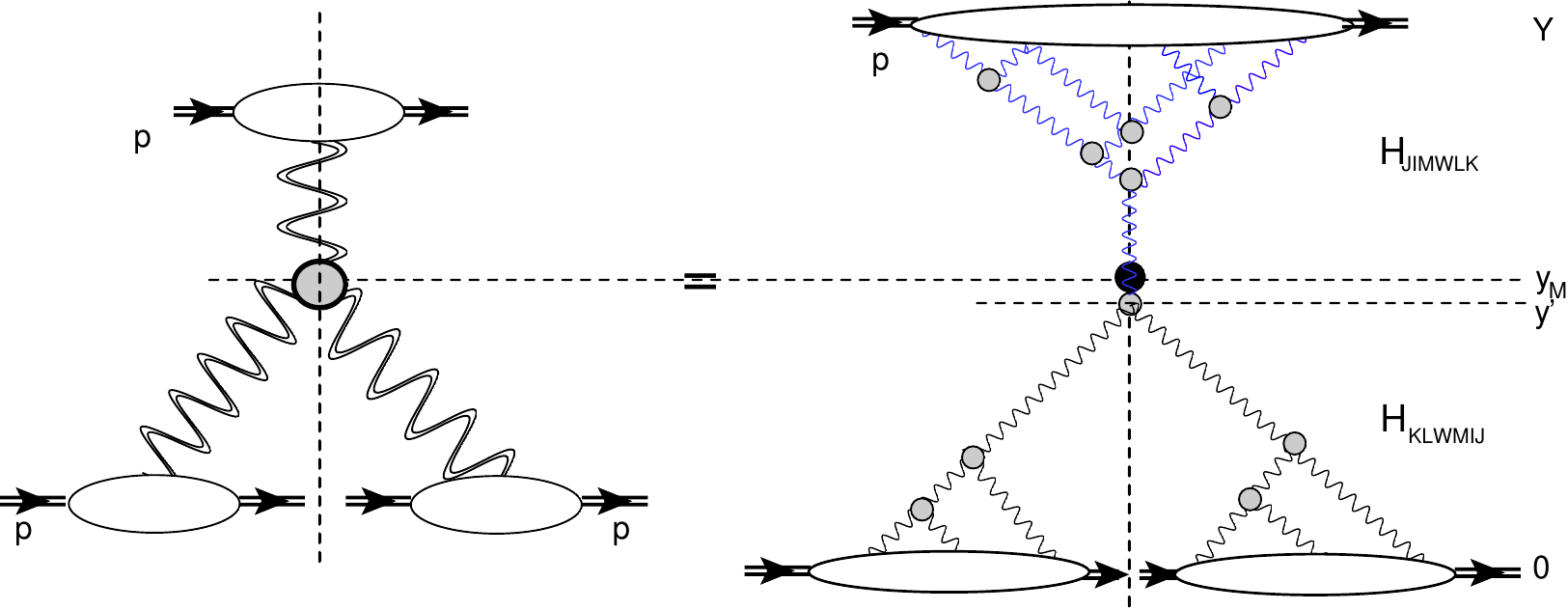} 
     \end{center}    
      \caption{The contribution of the dressed Pomeron to the diffraction production of   large mass in the  proton-proton scattering.
     The Balitsky-Kovchegov cascades  are described by $H_{\mbox{\tiny JIMWLK}}$ and by 
     $H_{\mbox{\tiny KLWMIJ }}$ (see Ref.~\cite{AKLL}).     The wavy lines denote the BFKL Pomerons. The gray circles are the triple Pomeron vertex while
the black circle denotes  $ \frac{1}{4 \pi^2} \int d^2 r_i\,d^2 b_i \frac{1}{r^4_i}$. The double wavy line describes the dressed Pomeron. $Y - y_M = \ln M^2$, where $M$ is the mass of produced hadrons. The vertical dashed line denotes the cut Pomeron.}
\label{ddpom}
   \end{figure}

In this section  we are going to study the cross section of the single diffractive dissociation. The physical picture of the process we are going to consider is the following: in DIS the virtual photon interacts with the hadron or nucleus breaking up into hadrons and jets in the final state. At the same time the target hadron (nucleus) remains intact. The particles produced as a result of  the hadron breakup do not fill the whole rapidity interval, leaving a rapidity gap between the target and the  slowest produced particle as a function of the invariant mass of the produced hadrons $M$.
 The diffractive production of   hadrons with large mass is  intimately    related to the triple Pomeron diagram which is shown in \fig{ddpom}. The three Pomeron vertex can be found from the Balitsky-Kovchegov~\cite{BK}  non-linear equation:

  \bea \label{BK}
&&\frac{\partial}{\partial Y}N\Lb \vec{x}_{10}, \vec{b} ,  Y; R \Rb = \nn\\
&&\bas\!\! \int \frac{d^2 \vec{x}_2}{2\,\pi}\,K\Lb \vec{x}_{02}, \vec{x}_{12}; \vec{x}_{10}\Rb \Bigg(N\Lb \vec{x}_{12},\vec{b} - \h \vec{x}_{20}, Y; R\Rb + 
N\Lb \vec{x}_{20},\vec{b} - \h \vec{x}_{12}, Y; R\Rb - N\Lb \vec{x}_{10},\vec{b},Y;R \Rb\nn\\
&&-\,\, N\Lb \vec{x}_{12},\vec{b} - \h \vec{x}_{20}, Y; R\Rb\,N\Lb \vec{x}_{20},\vec{b} - \h \vec{x}_{12}, Y; R\Rb\Bigg)
\eea
where $\vec{x}_{i k}\,\,=\,\,\vec{x}_i \,-\,\vec{x}_k$  and $ \vec{x}_{10}
 \equiv\,\vec{r}$, $\vec{x}_{20}\,\equiv\,\vec{r}' $ and $\vec{x}_{12}
 \,\equiv\,\vec{r}\,-\,\vec{r}'$.  $Y$ is the rapidity of the scattering
 dipole and $\vec{b}$ is the impact factor. $K\Lb \vec{x}_{02}, \vec{x}_{12};
 \vec{x}_{10}\Rb$ is the kernel of the BFKL equation which has the following form:
 \beq \label{KER}
 K\Lb \vec{x}_{02}, \vec{x}_{12};\vec{x}_{10}\Rb \,\,\,=\,\,\,\frac{x^2_{10}}{x^2_{12}\,x^2_{02}}  \eeq

     The last term of \eq{BK} gives the triple Pomeron contribution. Using \eq{ME} we can re-write the equation given by \fig{ddpom} in the following analytical form:
     \bea \label{DDME}
  \frac{d \sigma_{sd}\Lb Y, y_M \Rb}{d y_M}\,\,&=&\,\,\frac{2}{4\,\pi^2}\,\int \frac{d^2 r}{ r^2}  \,\int d^2 b\, N_p\Lb Y- y_M, \vec{r}, \vec{b } \Rb\,\\
  &\times& \Bigg\{ \,\bas\,\int d^2 b' \int \frac{d^2 r'}{2\,\pi} \frac{1}{r'^2} N_p\Lb y_M, \vec{r}',\vec{b}' - \h \Lb \vec{r} \,-\,\vec{r}'\Rb \Rb\,\frac{1}{\Lb \vec{r} \,-\,\vec{r}'\Rb^2}N_p\Lb y_M,  \vec{r} \,-\,\vec{r}',\vec{b}' - \h \vec{r}'\Rb \Bigg\}  \nn
  \eea   
    where $Y - y_M= \ln M^2$ where $M$ is the mass of produced hadrons (see \fig{ddpom}). 
       
       In \eq{DDME} the typical values of $r (r')$ are  $r \,\sim\,1/Q_s\Lb Y - Y_M,b\Rb$  and $r' \,\sim\,1/Q_s\Lb Y_M,b'\Rb $. For understanding the dependence on $y_M$ we can consider two different cases.
       \begin{enumerate}
       \item \quad $ Q_s\Lb Y - Y_M,b\Rb       \,\gg\,Q_s\Lb Y_M,b'\Rb$    
          
      In this case  
       we see that $r \,\ll\,r'$ and the integral over $r'$ takes the form:
       \beq \label{DD1}
 I\Lb r\Rb\,\equiv\,\,  \int\limits_{r' > r} \frac{d^2 r'}{2\,\pi} \frac{1}{r'^4} N\Lb y_M, \vec{r}',\vec{b}'\Rb\,N\Lb y_M,  \vec{r}',\vec{b}' \Rb \,\,\propto \,\,
       Q^2_s\Lb y_M, b'\Rb      
        \eeq  
        where we consider that $b \sim 1/\mu \,\gg\,r (r') \,\sim\,1/Q_s$.
        Hence, we  infer that the rapidity dependence of   $  \frac{d \sigma_{sd}\Lb Y, y_M \Rb}{d y_M} $ is $  \frac{d \sigma_{sd}\Lb Y, y_M \Rb}{d y_M} \,\,\propto\,\, \int d^2 b' \,Q^2_s\Lb y_M, b'\Rb $. However, it is not correct. Indeed, the integration over $r$ takes the form
    \beq \label{DD10}
\int \frac{d^2 r}{ r^2}  \,\, N_p\Lb Y- y_M, \vec{r}, \vec{b } \Rb I\Lb r\Rb
   \eeq     
    This integral converges at large $r$ only due to a decrease of function $I(r)$  which can occur only for $r \,>\,1/ Q_s\Lb Y_M,b\Rb $. Therefore, in the region of $1/ Q_s\Lb Y_M,b\Rb\,\,>\,r\,>\,\, 1/ Q_s\Lb Y - Y_M,b\Rb  $ we have a logorithmic integral which leads to the contribution:
     \beq \label{DD10}
\int \frac{d^2 r}{ r^2}  \,\, N_p\Lb Y- y_M, \vec{r}, \vec{b } \Rb I\Lb r\Rb =\,\,C_{r'> r} \,\,+\,\,\ln\Lb \frac{ Q_s\Lb Y - Y_M,b\Rb}{Q_s\Lb Y_M,b\Rb} \Rb\,\,=\,\,C_{r'> r}\,+\,\lambda \Lb Y \,-\,2\,y_M\Rb \eeq      
  Therefore, we expect that the contribution to the diffraction production from this kinematic region has a general form:
  \beq \label{DD12}
    \frac{d \sigma_{sd}\Lb Y, y_M; r', r  \Rb}{d y_M}  \propto Q^2_s\Lb y_M, b'\Rb \Big(C_{r'> r}\,+\,\lambda \Lb Y \,-\,2\,y_M\Rb\Big)
    \eeq      
      
 \item \quad          $ Q_s\Lb Y - y_M, b\Rb       \,\ll\,Q_s\Lb Y_M,b'\Rb$

 In this kinematic region the typical $r\,\,\gg\,\,r'$ and we obtain the integral over $r$ in the form:
         \beq \label{DD2}
       \int \frac{d^2 r}{2\,\pi} \frac{1}{r^4} N_p\Lb Y - y_M, \vec{r},\vec{b}\Rb\,N_p\Lb  y_M, \vec{r},\vec{b}'\Rb \,\,\propto \,\,
       Q^2_s\Lb Y - y_M, b\Rb      
        \eeq     
   leading to   $  \frac{d \sigma_{sd}\Lb Y, y_M \Rb}{d y_M} \,\,\propto\,\, \int d^2 b\, Q^2_s\Lb  Y -y_M, b\Rb $. Note, that $N\Lb  y_M, \vec{r},\vec{b}'\Rb\, \to\,1$  in this kinematic region.  Repeating the same estimates as in the case 1 for integration over $r'$ we conclude that
   
     \beq \label{DD21}
    \frac{d \sigma_{sd}\Lb Y, y_M; r',  r  \Rb}{d y_M}  \propto Q^2_s\Lb Y -  y_M, b'\Rb \Big(C_{r > r'}\,-\,\lambda \Lb Y \,-\,2\,y_M\Rb\Big)
    \eeq       
    From \eq{DD12} and \eq{DD21} we conclude that $  
    \frac{d \sigma_{sd}\Lb Y, y_M; r', r  \Rb}{d y_M} $ has maximum in the region of $y_M \approx \h Y$. It is easy to see that $C_{r > r'} \,>\,C_{r' > r}$ and hence 
   the maximum is shifted to $y_M > \h Y$. 
    
   \end{enumerate}
     \begin{figure}[ht]
     \begin{center}
     \includegraphics[width=\textwidth]{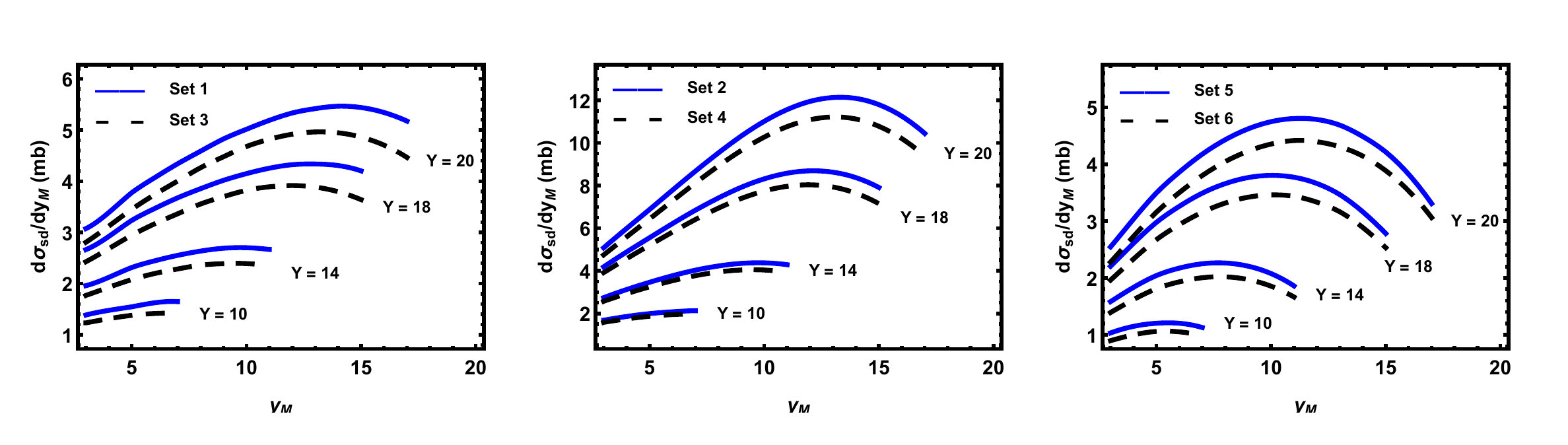} 
     \end{center}    
      \caption{Cross section of the single diffraction production $\frac{d \sigma_{sd}\Lb Y, y_M \Rb}{d y_M } $ versus $y_M$ at different values of $y$ for sets 1-6 of Ref.~\cite{CLS}. 
       }
\label{sdvsym}
   \end{figure}

     \begin{figure}[ht]
     \begin{center}
     \includegraphics[width=\textwidth]{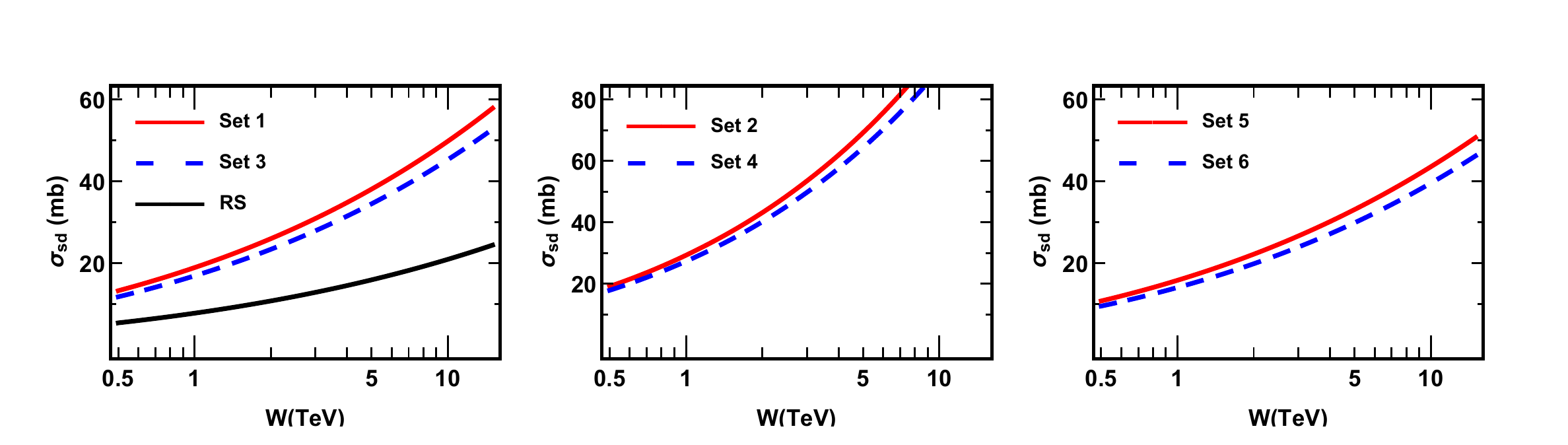} 
     \end{center}    
      \caption{Cross section of the single diffraction production $\int d y_m\,\frac{d \sigma_{sd}\Lb Y, y_M \Rb}{d y_M } = \sigma^{diff}$ versus $Y$  for different  sets of Ref.~\cite{CLS}. The solid black line, which is denoted by RS,  corresponds to the saturation model of Ref.~\cite{SATMOD14}. }
\label{sdvsy}
   \end{figure}

Hence, we can expect that

\beq \label{DD20}
\sigma^{diff}\,\,=\,\,\int dy_M \,  \frac{d \sigma_{sd}\Lb Y, y_M\Rb}{d y_M}\,\,\propto\,\,\int d^2 b\, Q^2_s\Lb \h Y, b\Rb\,\Big( {\rm Const} \,\,+\,\,\lambda \,Y\Big)\eeq

In \fig{sdvsym} we plot the estimates  for $\frac{d \sigma_{sd}\Lb Y, y_M \Rb}{d y_M } $
in different parameterizations of Ref.~\cite{CLS}. 
   At not very large $Y$ the cross section increases with the increase of rapidity gap ($ y_{\rm gap} = Y - y_M$), which agrees with the result of the traditional triple pomeron description of the diffractive dissociation. However, as $y_{\rm gap}$ gets very high  and reaches   the values of rapidity at saturation,   the cross section reaches a maximum and starts decreasing.
One can see that distribution over $y_M$ has a maximum in the region of $y_m \approx \h Y$,  which has been expected from the qualitative discussions above. It should be mentioned that such a maximum follows from the non linear evolution equation for diffractive dissociation processes in QCD~\cite{KOLE}.

  In \fig{sdvsy} the values of $\sigma^{diff}  \,\,=\,\,\int^{Y - y_0}_{y_0} d y_M\,\frac{d \sigma_{sd}\Lb Y, y_M \Rb}{d y_M }$ are plotted.  The value of $y_0$ is chosen $y_0=3$, which reflects our belief  that  we can consider the Pomeron exchange starting with rapidity $  \geq\,y_0$. One can see that the $Y$ dependence in this figure reproduces the estimates of \eq{DD20}. On the other hand, the values turn out to be very large and, hence, the shadowing corrections are needed.

\begin{figure}[ht]
     \begin{center}
     \includegraphics[width=\textwidth]{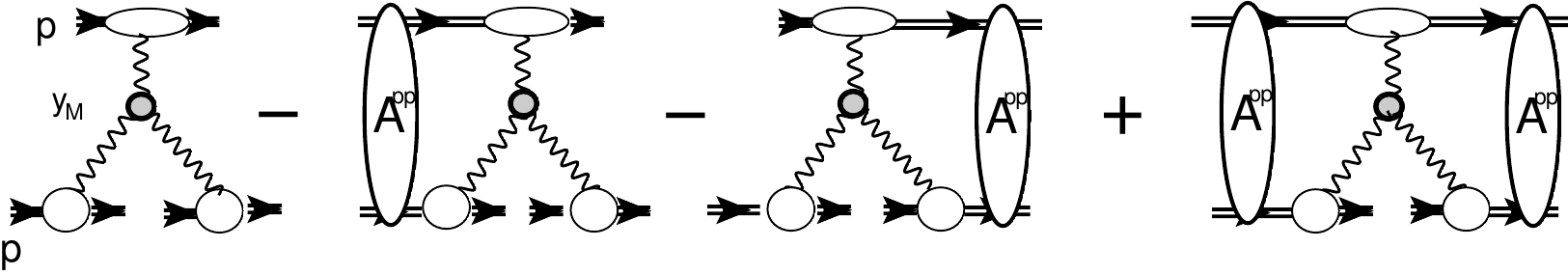} 
     \end{center}    
      \caption{ Shadowing correction to the single diffraction production. $A^{pp}\Lb Y, b\Rb$  is given by \eq{EIK}.}
\label{sdsc}
   \end{figure}

  In \fig{sdsc} it shown the typical eikonal type shadowing  corrections.~\cite{GOLED}, which suppress the large values of the diffraction cross section.
 The sum of the diagrams in \fig{sdsc} results in the following formula for $ \frac{d \sigma_{sd}\Lb Y, y_M \Rb}{d y_M} $:
\beq \label{DD3}
 \frac{d \sigma_{sd}\Lb Y, y_M \Rb}{d y_M} \,\,=\,\,\int d^2 b \,\, \,e^{ - 2\, N^p_p\Lb Y, b\Rb }\,\frac{d \sigma_{sd}\Lb Y, y_M; b;\eq{DD4}\Rb}{d y_M\, d^2 b}\eeq
where
\bea \label{DD4}
\frac{d \sigma_{sd}\Lb Y, y_M; b\Rb}{d y_M \,d^2 b}\,\,&=&\,\,\frac{2}{4\,\pi^2}\,\int d^2 b' \int \frac{d^2 r}{ r^2}  \, N_p\Lb Y- y_M, \vec{r}, \vec{b} \,-\,\vec{b}'\Rb\,\\
  &\times& \Bigg\{ \,\bas\,\int \frac{d^2 r'}{2\,\pi} \frac{1}{r'^2} N_p\Lb y_M, \vec{r}',\vec{b}' - \h \Lb \vec{r} \,-\,\vec{r}'\Rb \Rb\,\frac{1}{\Lb \vec{r} \,-\,\vec{r}'\Rb^2}N_p\Lb y_M,  \vec{r} \,-\,\vec{r}',\vec{b}' - \h \vec{r}'\Rb \Bigg\}  \nn
  \eea

It should be noted that the shadowing corrections in \fig{sdsc} stems from the simple eikonal model as well as \eq{EIK},  and could be used only to show the scale of the effect. In particular, for $W=13\,TeV$ we evaluated  $\sigma_{\rm s.d.} = 4 \,mb$  indicating that the shadowing correction can be large.

\begin{figure}[ht]
     \begin{center}
     \includegraphics[width=0.7\textwidth]{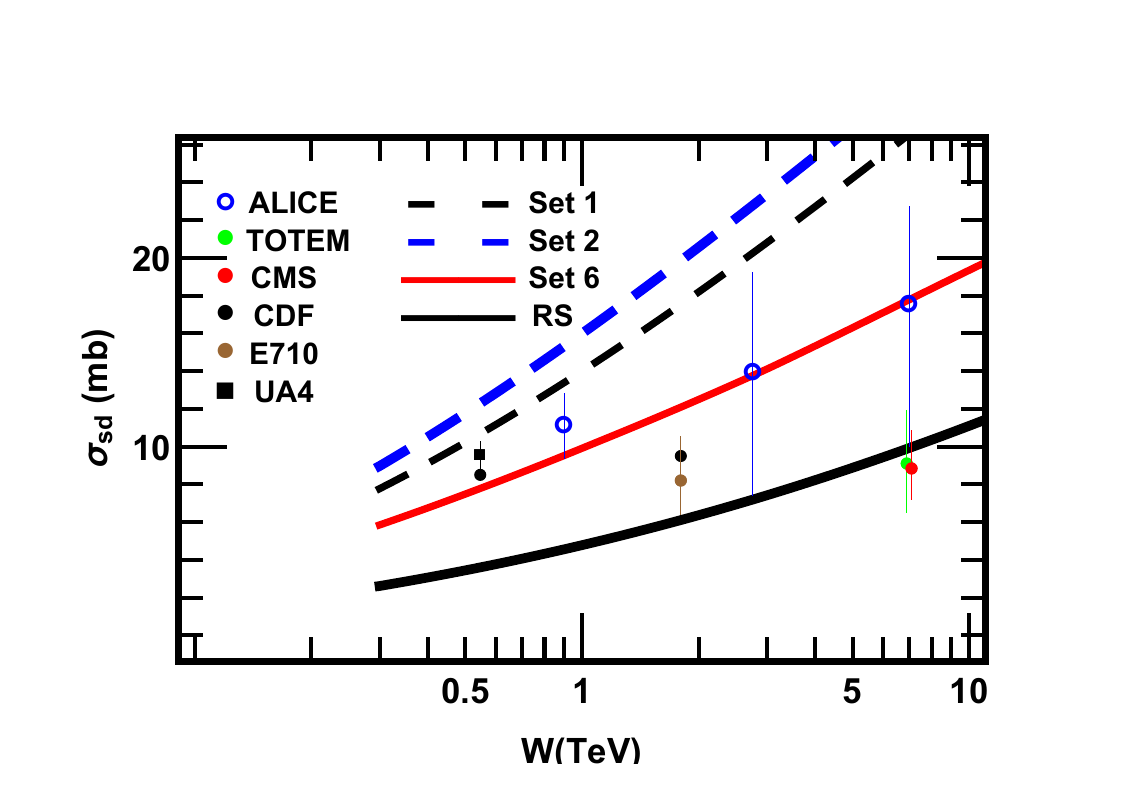} 
     \end{center}    
      \caption{ The estimates for the single diffractive cross section using the simple eikonal formula of \eq{DD3}
      for sets of parameters in Ref.~\cite{CLS} and for  saturation model of Ref.~\cite{SATMOD14}(RS).
      The data are taken from Refs.~\cite{PDG,JKAS,ALICE1,CMS1} .}
\label{sdsc1}
   \end{figure}

The values for the cross section of diffractive production with the  simplified shadowing correction of \eq{DD3} are plotted in \fig{sdsc1}.  One can see that the shadowing is very important,  but the simple  eikonal formula  cannot pretend to take them all into account on the theoretical grounds. Again as for $\sigma_{tot},\sigma_{el}$ and $B_{el}$, we see the need for the theoretical approach for the shadowing corrections. The experience with the simple models~\cite{MUDI,LELU1,LELU2,MUSA,BIT,AKLL} shows   the eikonal formula can be used only as a rough estimate.

\section{Conclusions}
      In this paper         we suggested a new approach to the structure of the soft Pomeron,  based on the $t$-channel unitarity: we expressed the exchange of the soft Pomeron through the interaction of the dipole of  small size on the order of $1/Q_s(Y)$  
  ($Q_s(Y)$ is the saturation momentum)  with the hadrons. 
 Thereby, it is shown that the typical distances  in so called soft processes turns out to be small $r \sim 1/Q_s\Lb \h Y \Rb $, where  $Y \,=\,\ln s$.  This fact opens new possibilities for describing the soft interactions in the framework of the Colour Glass Condensate(CGC) approach, putting the high energy phenomenology on  solid theoretical basis.

 The energy dependence of the scattering amplitude  due to Pomeron exchange is determined by the saturation momentum  $ N^p_p\Lb \pom\Rb \,\propto\,Q^2_s\Lb \h Y\Rb$ (see \eq{ME}), which increases as a power of energy.  Therefore, the suggested Pomeron leads to the violation of the Froissart theorem, but
 $ N^p_p\Lb \pom\Rb \,\propto\,Q^2_s\Lb \h Y\Rb \,\,\propto\,e^{\lambda\,Y} $ with $\lambda  \approx\,0.1 - 0.13$ is in perfect agreement with phenomenological Donnachie-Landshoff 
 Pomeron~\cite{DOLA}. We believe, that our approach can be the good first approximation to start discussion of the soft process in  CGC approach.
 
 We made an attempt to describe the value of the Pomeron exchange directly from our knowledge of the deep inelastic processes.  
 First off,    we have to mention that we  cannot describe DIS processes in framework of CGC in spite of the well known Balitsky-Kovchegov evolution equation. As we have discussed, B-K approach suffers from unsolved difficulties, including  the large impact parameter ($b$)  behaviour that violates the Froissart theorem~\cite{KW1,KW2,KW3}. We have to use models which introduce to B-K equation an additional exponential decrease at large $b$. Second, the models have been checked against the experimental data on DIS. However, the  energy range of the experimental data are quite different from the one of soft interaction.  The lesson, that we learned, is that  some sets of parameterizations,  which describe the DIS, lead to reasonable description of the soft high scattering but other sets cannot be described. In spite of the large dispersion of the estimates we see several general features, which could be useful in further development of the CGC approach to soft interaction. First, almost in all estimates we need strong shadowing corrections, both to obtain the reasonable values of the experimental observable, and to describe the shrinkage of the diffraction peak. The dressed Pomeron that we have introduced can not describe this shrinkage even on qualitative level.  Second, the best description  $\sigma_{tot},\sigma_{el},B_{el}$ and $\sigma_{diff}$ we obtain from the impact parameter dependence that
  incorporates in the BFKL equation the
 Gribov's   diffusion (see Refs.~\cite{LEPION,GOLEB}  and references therein).

     However, we can look on our attempts to obtain the soft Pomeron from the DIS saturation approach at a different angle, stating that sets 5 and 6 of Ref.~\cite{CLS}  are good candidates  for the global fit of DIS and soft interaction experimental data at high energies. The possibility of such combined description is both encouraging and exiting.

     The approach, that we developed here, was started in Refs.~\cite{GLR,MUDI}    for the exchange of the BFKL Pomeron, and it is the modified version of the MPSI treatment~\cite{MUPA,MPSI}, in which we use the properties of the BFKL Pomeron to absorb the QCD Born amplitude  in the closed expression. 
     
\section{Acknowledgements}
   We thank our colleagues at Tel Aviv university and UTFSM for
 encouraging discussions. Part of this work has been done, while one of us (M.S.) has been visiting the  Ohio State University, and he  
 wishes to express his deep gratitude to Prof. Yu. Kovchegov  for  hospitality and help during this visit. This research was supported  by  ANID PIA/APOYO AFB180002 (Chile) and  Fondecyt (Chile) grants  
 1180118 and 1191434, Conicyt Becas (Chile) and PIIC 009/2021, DPP, Universidad Tecnica Federico Santa Maria

\end{document}